\documentclass[aps,prc,superscriptaddress]{revtex4-1}
\usepackage{graphicx}
\usepackage{multirow}
\usepackage{color}
\usepackage{amssymb}
\usepackage{bbm}
\usepackage{amsmath}
\usepackage{color}
\usepackage{lineno}
\usepackage{enumitem}
\usepackage{scalerel,stackengine}
\usepackage{xcolor}
\usepackage{comment}
\usepackage[normalem]{ulem}

\renewcommand{\Im}{\mathop{\mathrm{Im}}}

\def\nuc#1#2{\relax\ifmmode{}^{#1}{\protect\text{#2}}\else${}^{#1}$#2\fi}

\newcommand{\be}{\begin{eqnarray}}
\newcommand{\ee}{\end{eqnarray}}

\newcommand{\bwt}{\begin{widetext}}
\newcommand{\ewt}{\end{widetext}}

\stackMath
\newcommand\reallywidehat[1]{%
\savestack{\tmpbox}{\stretchto{%
  \scaleto{%
    \scalerel*[\widthof{\ensuremath{#1}}]{\kern-.6pt\bigwedge\kern-.6pt}%
    {\rule[-\textheight/2]{1ex}{\textheight}}%WIDTH-LIMITED BIG WEDGE
  }{\textheight}%
}{0.5ex}}%
\stackon[1pt]{#1}{\tmpbox}%
}

\begin{document}
%\begin{CJK*}{GB}{song}
%\preprint{}

\title{Advancing the IAV Model with CDCC Wave Functions for Realistic Descriptions of Two-Body Projectile Breakup}


\author{Jin Lei}
\email[]{jinl@tongji.edu.cn}
%\thanks{}

%\altaffiliation{Present address: Institute of Nuclear and Particle Physics, and Department of Physics and Astronomy, Ohio University, Athens, Ohio 45701, USA}
%\thanks{}
\affiliation{School of Physics Science and Engineering, Tongji University, Shanghai 200092, China.}
% \affiliation{Institute for Advanced Study of Tongji University, Shanghai 200092, China}

\author{Antonio M. Moro}
\email[]{moro@us.es}
%\thanks{}

\affiliation{Departamento de FAMN, Universidad de Sevilla,
Apartado 1065, 41080 Sevilla, Spain.}


\begin{abstract}
Inclusive breakup is an important reaction mechanism of reactions induced by weakly bound nuclei. The Ichimura, Austern, and Vincent (IAV) model is widely used to analyze inclusive breakup processes and is based on a Distorted Wave Born Approximation (DWBA). However, the validity of the DWBA form for inclusive breakup requires further exploration. In this study, we present a derivation of the IAV model, using the continuum-discretized coupled-channels (CDCC) wave function, and apply it to the $d+^{93}$Nb reaction. We examine the differences between the CDCC-IAV and DWBA-IAV models by artificially modifying the binding energy of the deuteron using two distinct types of optical potential. Our findings indicate that the CDCC method potentially provides a more fundamental description of the $d$+$A$ interaction and may provide a more realistic description of the interior part of the wave function. This study offers significant potential in increasing the precision and dependability  of weakly bound nuclei inclusive breakup process estimation within a fully quantum mechanical model.
\end{abstract}

% 25.70.Mn, Projectile and target fragmentation
% 24.10.Eq 	Coupled-channel and distorted-wave models
% 25.45.-z  2H-induced reactions
% 24.87.+y 	Surrogate reactionsç

%\keywords{}
\date{\today}%
\maketitle

\section{Introduction \label{sec:intro}}
Nuclear reactions involving weakly bound nuclei have attracted much attention in recent years due to their unique and often exotic properties~\cite{Duan2022,Pietro19,Pes17,Duan20,Yang18,Wang21}. Because of their weak binding these nuclei usually behave differently from more tightly bound nuclei. Thus, reactions of weakly bound nuclei can lead to intriguing and occasionally surprising phenomena~\cite{JHA20201}. For example, the breakup of the weakly bound nucleus into smaller fragments can provide insight into the structure of the nucleus and the behavior of nuclear forces at low energies. The study of reations induced by weakly bound nuclei plays a crucial role in advancing our understanding of fundamental nuclear physics.

An important reaction mechanism of reactions induced by weakly bound nuclei is the inclusive breakup, expressed as $a (=b +x) + A \rightarrow b + B^*$. Here, $a = b + x$ and $B^*$ represents any possible state between the $x$ and $A$ systems, including elastic breakup (EBU). In EBU, $x$ undergoes elastic scattering with $A$ and both the fragments and the target are left in their respective ground states. Various nonelastic breakup (NEB) channels are also possible, including the exchange of nucleons between $x$ and $A$, projectile dissociation accompanied by target excitation, and fusion of $x$ by $A$, among others.

The Ichimura, Austern, and Vincent (IAV) model~\cite{IAV85}, proposed during the 1980s, allows for the treatment of inclusive breakup processes. The first version of the model was based on the Distorted Wave Born Approximation (DWBA) so that the entrance channel wavefunction was described using a distorted wave generated with an appropriate optical model potential. The model was later extended to incorporate a three-body description of the entrance channel by means of the continuum-discretized coupled-channels (CDCC) wave function~\cite{Austern87},  an approximation of the exact three-body wave function, solution of the Faddeev equations~\cite{HUSSEIN1990269}. Despite the Faddeev equation being the most accurate method, obtaining an exact solution can be computationally challenging for a scattering problem. As a result, many applications in the literature utilize a simpler solution by directly calculating EBU cross sections with the CDCC method and solving NEB processes using the DWBA framework~\cite{Jin15,Jin15b,Jin17,Jin18,Jin18b,Jin19}. However, it should be noted that while the original IAV model utilizes the lowest order (elastic channel) wave function in DWBA form, the validity of the DWBA form for the inclusive breakup process requires further exploration.

In our previous work~\cite{Jin19}, we proposed the first implementation of the CDCC wave function to the IAV model. We demonstrated that the DWBA form can approximate the CDCC wave function for both the deuteron and $^6$Li cases, at least in the reactions analysed in that work. The CDCC wave function naturally separates the contributions between bound states and continuum states, enabling the study of continuum effects in nonelastic breakup processes. In other words, we can examine nonelastic breakup as a two-step process such as $a + A \rightarrow b + x + A \rightarrow b + B^*$, or as a one-step process such as $a + A \rightarrow b + B^*$. We can verify the latter by comparing the results of applying the full CDCC wave function to the IAV model with only the ground state portion of the full CDCC wave function.

The former study has direct implications in the physical interpretation of incomplete fusion (ICF), a type of nuclear reaction between a projectile and a target nucleus in which  a fraction of the projectile nucleus fuses with the target nucleus and the remaining part moves forward without fusing. By definition, ICF is part of the NEB process. Several models~\cite{Ferreira23,Diaz07,Marta14,Kolinger18} exist for calculating ICF cross-sections, but the challenging problem remains how to provide an unambiguous calculation within a fully quantum mechanical model. Previous literature describes the ICF process as a breakup fusion, a two-step reaction where the projectile first breaks up into $b$ and $x$, and the fragment $x$ is subsequently absorbed by the target nucleus. However, recent experimental results question the reliability of the breakup fusion model, suggesting instead that ICF products are consistent with a direct, one-step mechanism~\cite{Cook19}. Our prior work identifies that the ICF entails a combination of one-step and two-step process; however, it is predominantly dominated by the one-step process.

In this paper, we provide a rigorous derivation of the IAV model using the CDCC wave function and apply it to the d+$^{93}$Nb reaction.  Additionally, we implement a toy model to investigate the effects of varying the binding energy of the deuteron, comparing the NEB cross sections obtained via the full CDCC wave function (CDCC-NEB) and the DWBA (DWBA-NEB). Our approach offers significant potential to enhance the precision and dependability of weakly bound nuclei inclusive breakup process estimation within a fully quantum mechanical model.

The structure of the paper is as follows. Section~\ref{sec:II} provides a thorough derivation of the IAV model with CDCC wave function. In Sec.~\ref{sec:III}, we apply the formalism to inclusive reactions induced by deuterons. Lastly, in Sec.~\ref{sec:IV}, we summarize the main findings of this study and provide an overview of potential future developments.

\section{Theoretical framework}
\label{sec:II}

%-------------------------------------
\subsection{The CDCC wave function}
%-------------------------------------
The CDCC technique was developed as an approximation to the Faddeev equations~\cite{Raw74,Yah86,Austern87,Austern89,Faddeev}. Its main objective is to provide a reliable and practical approach to describe reactions that involve three-body breakup.
For example, consider a reaction system $a+A$ where $a=b+x$, such as $d=n+p$ or $^6$Li$=\alpha +d$. In this case, the effective three-body Hamiltonian may be expressed as follows:
\begin{equation}
\label{eq:H}
H = H_\mathrm{proj} + T_{R} + U_{bA} + U_{xA},
\end{equation}
where $H_\mathrm{proj}$ represents the projectile internal Hamiltonian, i.e., $H_\mathrm{proj}= T_r+V_{bx}$, with $T_r$  the kinetic energy operator, and $V_{bx}$  the binding potential of projectile. The operators $U_{bA}$ and $U_{xA}$ describe the optical potentials for the elastic scattering of corresponding sub-systems $b+A$ and $x+A$ at the same energy per nucleon of the incident projectile. Thus, from the perspective of the internal system of a two-body projectile, the unit operator of this three-body model space can be expressed as:
\begin{equation}
\label{eq:unit}
\mathbbm{1} = \sum_{\alpha  n_0}\int R^2 dR | \phi^{n_0}_{bx}R\alpha \rangle \langle \phi^{n_0}_{bx}R\alpha |
+ (2\pi)^{-3}\sum_{\alpha }\int R^2 dR \int d\vec{k} | \phi^{\vec{k}(+)}_{bx}R\alpha \rangle \langle \phi^{\vec{k}(+)}_{bx}R\alpha |,
\end{equation}
where the index $n_0$ runs over the discrete spectrum, the integral $d\vec{k}$ is over the 
continuous spectrum of $H_\mathrm{proj}$, $R$ is the relative coordinate between $a$ and $A$, and $|\alpha \rangle$ is the angular momentum eigenstates with 
\begin{equation}
| \alpha  \rangle = | (l_a (s_b s_x)s_{bx})j_{bx} (\lambda_A s_A)j_A; J  M \rangle, 
\end{equation} 
where $s_b$, $s_x$ and $s_A$ are the spins of $b$, $x$, and $A$ particles, respectively, while $l_a$ and $\lambda_A$ correspond, respectively, to the orbital angular momentum of the pair $b$-$x$, and the system $a$-$A$. The $J$ and $M$ values represent the total angular momentum and its projection onto the z-axis. For practical reasons, the integral over projectile scattering states in 
Eq.~(\ref{eq:unit}) is discretized and truncated at a maximum energy. Several methods can be used for discretization;  for this study, a well-known bin method was chosen. As per the definition given in Ref.~\cite{thompson_nunes_2009}, the radial functions of the continuum bins are a superposition of the scattering eigenstates,
\begin{equation}
\langle r\beta  | \phi_{bx}^n \rangle = \frac{1}{r}\sqrt{\frac{2}{\pi N_k}} \int_{k_n}^{k_{n+1}} g(k) f^{j_{bx}}_{l_a} (kr) dk,  
\end{equation}
where $|\beta  \rangle =  | l_a (s_b s_x)s_{bx};j_{bx} \rangle$, $f^{j_{bx}}_{l_a}(kr)$ is the radial part of scattering wave function at momentum $\hbar \vec{k}$, $g(k)$ is the weight function, and the normalization constant is chosen as $N_k=\int_{k_n}^{k_{n+1}} |g(k)|^2 dk$. One should note that the bin state wave functions do not depend on the direction of $\vec{k}$. 
For the continuum part of Eq.~(\ref{eq:unit}), we can separate it into different cells and insert the bin state into the corresponding integration cell, resulting,
\begin{align}
& (2\pi)^{-3} \sum_{\alpha  } \sum_{n} \int R^2 dR \int_{k_n}^{k_{n+1}} d \vec{k} | \phi_{bx}^{\vec{k}(+)}R\alpha \rangle \langle \phi_{bx}^{\vec{k}(+)}R\alpha  | \nonumber\\
&= (2\pi)^{-3} \sum_{\alpha  } \sum_{n} \int R^2 dR \int_{k_n}^{k_{n+1}} d \vec{k} | \phi_{bx}^n R\alpha  \rangle \langle  \phi_{bx}^{n\alpha} | \phi_{bx}^{\vec{k}(+)}\rangle \langle \phi_{bx}^{\vec{k}(+)}|\phi_{bx}^{n\alpha} \rangle \langle  \phi_{bx}^n R\alpha |  \nonumber \\
& = \sum_{\tilde{\alpha} } \int R^2 dR |\phi^n_{bx} R\alpha \rangle \langle \phi^n_{bx} R\alpha|, 
\end{align}
%where one use $\tilde{\alpha}$ to replace $\alpha$ and $n$ index and
where in the last line we have introduced the notation $\tilde{\alpha}\equiv \{\alpha, n\}$ and
\begin{align}
\langle  \phi_{bx}^{n\alpha} | \phi_{bx}^{\vec{k}(+)}\rangle = \frac{1}{\sqrt{2j_{bx}+1}}\sqrt{\frac{2}{\pi N_k}} \frac{2\pi^2}{k} g^*(k)e^{i \sigma_{la}}\sum_{m_{j_{bx}}}\sum_{m_{l_a}}\sum_{m_{s_{bx}}}\langle l_a m_{l_a} s_{bx} m_{s_{bx}} | j_{bx} m_{j_{bx}} \rangle i^{l_a} Y_{l_a}^{m_{l_a}*}(\hat{k}) \chi_{s_{bx}}^{m_{s_{bx}}}(\vec{\sigma}_{bx}), 
\end{align}
where $\sigma_{l_a}$ are the Coulomb phase shifts and  $\chi_{s_{bx}}^{m_{s_{bx}}}(\vec{\sigma})$ are the normalized spin functions.
The indexes $n_0$ and $n$ can be combined into a single index $n$. For this study, $n\leq 0$ denotes the bound state while $n>0$ indicates the discretized continuum states. By doing so, the unit operator in Eq.~(\ref{eq:unit}) can be simplified.
\begin{equation}
\mathbbm{1} \approx \sum_{\tilde{\alpha}} \int R^2 dR |\phi^n_{bx} R\alpha \rangle \langle \phi^n_{bx} R\alpha| .
\end{equation}

For a reaction of the form $a(=b+x) + A$, the  initial state is characterized by
$j_a$, $m_a$, and $m_{A}$, where $j_a$, $m_a$, and $m_{A}$
are the spin of projectile and the magnetic quantum numbers of projectile and target, respectively. 
To simplify the notation, one can label these states as $j_\mathrm{in}=\{j_a,m_a,m_{A}\}$. 
The CDCC wave functions for given incoming states are the solution of Schr\"{o}dinger equation 
\begin{equation}
\label{eq:sch}
(E-H_\mathrm{proj}-T_R-U_{bA}-U_{xA})\Psi^{j_\mathrm{in}(+)} = 0.
\end{equation}
%which is conveniently rewritten as 
%\begin{equation}
%\label{eq:sch}
%(E-H_\mathrm{proj}-T_{R}) \Psi^{j_\mathrm{in}(+)}   = ( U_{bA} + U_{xA}) \Psi^{j_\mathrm{in}(+)}. 
%\end{equation}
%\amm{[I think eq. (9) could be omitted since it adds no additional information with respect to the previous one]} \jl{[ok we can remove this]}

In the CDCC method one solves the above equation by projecting it onto a bin state of $|\phi^n_{bx} R \alpha \rangle$,  so that Eq.~(\ref{eq:sch}) becomes
\begin{equation}
\label{eq:project}
\Big(E-\varepsilon_{n}-T_{aA} (R\alpha) \Big) \langle \phi^n_{bx}R\alpha |\Psi^{j_\mathrm{in}(+)}  \rangle 
= \langle  \phi^n_{bx} R\alpha | U_{bA}+U_{xA}| \Psi^{j_\mathrm{in}(+)}\rangle ,
\end{equation}
where 
\begin{equation}
T_{aA} (R\alpha) = -\frac{\hbar^2}{2\mu_{aA}} \Big[\frac{1}{R}\frac{\partial^2}{\partial R^2}R-\frac{\lambda(\lambda+1)}{R^2}\Big],
\end{equation}
$\varepsilon_{n}$ is calculated as $\langle \phi_{bx}^n |H_\mathrm{proj}| \phi_{bx}^n \rangle$, and $\mu_{aA}$ denotes the reduced mass of the $a-A$ system.
Inserting the unit operator in Eq.~(\ref{eq:project}), the right hand side becomes
\begin{equation}
\mathrm{R.H.S.}=\sum_{\tilde{\alpha}' }\int R'^2dR'\langle \phi^n_{bx} R\alpha |U_{bA}+U_{xA}| \phi^{n'}_{bx}R'\alpha' \rangle \langle  \phi^{n'}_{bx}R'\alpha' |\Psi^{j_\mathrm{in}(+)}\rangle. 
\end{equation} 
By assuming all the interactions are local, then the Schr\"{o}dinger equation becomes
\begin{equation}
\label{eq:CDCC}
\Big( E- \varepsilon_{n}- T_{aA} (R\alpha) \Big)\langle \phi^n_{bx} R\alpha| \Psi^{j_\mathrm{in}(+)} \rangle -
\sum_{\tilde{\alpha}' } U_{\tilde{\alpha}, \tilde{\alpha}'}(R) \langle \phi^{n'}_{bx} R\alpha'| \Psi^{j_\mathrm{in}(+)} \rangle=0.
\end{equation}
with the coupling potentials 
\begin{align}
U_{\tilde{\alpha}, \tilde{\alpha}'}(R) &  = \langle \phi_{bx}^n R \alpha  |U_{bA} + U_{xA}| \phi_{bx}^{n'}R\alpha' \rangle  \nonumber \\
& = \int d\vec{r} d\vec{R'}  d\vec{r'} d\vec{R''} \,  \langle  \phi^n_{bx} R \alpha| \vec{r} \vec{R'}\rangle
\langle \vec{r}\vec{R'}| U_{bA} + U_{xA} | \vec{r'}\vec{R''} 
\rangle \langle \vec{r'}\vec{R''}|\phi^{n'}_{bx} R \alpha' \rangle. 
\end{align}

As we have assumed the optical potentials to be local and independent on the internal degrees of freedom of the three fragments, and denoting $U=U_{bA} + U_{xA}$ we have, 
\begin{equation}
\label{eq:Upot}
U_{\tilde{\alpha}, \tilde{\alpha}'}(R) = \int d\vec{r} d\vec{R'}  
U(\vec{r},\vec{R'}) \langle \phi^n_{bx}R\alpha| \vec{r} \vec{R'}  \rangle 
\langle \vec{r} \vec{R'}  | \phi_{bx}^{n'}R\alpha'\rangle .
\end{equation}
In the above equation, $\langle \vec{r} \vec{R'}  | \phi_{bx}^{n'}R\alpha'\rangle$
takes the form
\begin{align}
\label{eq:phi}
\langle \vec{r} \vec{R'} | \phi_{bx}^{n'} R \alpha'  \rangle 
& = \sum_{L'S'}[(2L'+1) (2S'+1) (2j'_{bx}+1) (2j'_A+1) ]^{1/2}\left\{ \begin{array}{ccc}   l'_{a}  & \lambda'_{A} & L  \cr
s'_{bx} & s_{A}  & S \cr
j'_{bx} & j'_A & J
\end{array}\right\}  \langle \vec{r} \vec{R'} | rR(l'_a \lambda'_A)L ((s_bs_x)s'_{bx}s_A)S; JM \rangle \nonumber \\
& \times \langle r \beta' | \phi_{bx}^{n'}  \rangle
\end{align}
and likewise for $\langle \phi^n_{bx}R\alpha| \vec{r} \vec{R'}  \rangle $ so Eq.~(\ref{eq:Upot}) becomes 
\begin{align}
\label{eq:ucouple}
U_{\tilde{\alpha},\tilde{\alpha}' } (R) & = \sum_{LS} (2S+1) [(2j'_{bx}+1)(2j_{bx}+1)(2j'_A+1)(2j_A+1)]^{1/2}
\left\{ \begin{array}{ccc}   l'_{a}  & \lambda'_{A} & L  \cr
s_{bx} & s_{A}  & S \cr
j'_{bx} & j'_A & J
\end{array}\right\}
\left\{ \begin{array}{ccc}   l_{a}  & \lambda_{A} & L  \cr
s_{bx} & s_{A}  & S \cr
j_{bx} & j_A & J
\end{array}\right\} \ \nonumber \\
& \times \int r^2 dr  \langle r \beta' | \phi_{bx}^{n'}  \rangle \langle \phi_{bx}^{n}  |  r \beta   \rangle
\int d\Omega_r d\Omega_R \sum_{M_L} \sum_{m_l m'_l} \sum_{m_\lambda m'_\lambda} Y_{l_a'}^{m'_l} (\hat{r}) Y_{\lambda_A'}^{m'_\lambda} (\hat{R}) Y_{l_a}^{m_l *} (\hat{r}) Y_{\lambda_A }^{m_\lambda *} (\hat{R}) \nonumber \\
& \times \langle l_a' m'_l \lambda_A' m'_\lambda | L M_L \rangle \langle l_a m_l \lambda_A m_\lambda | L M_L \rangle  U(\vec{r},\vec{R}) .
\end{align}
In the above equation, the potential and spherical harmonics $Y$ depend on the angles 
$\hat{r}$ and $\hat{R}$. 
For the evaluation of the integral, we choose $\hat{R}$ as $z-$direction, and assume $\vec{r}$  is in the $x-y$ plane:
\begin{equation}
\vec R = \left( \begin{array}{c}
0 \cr 0 \cr R
\end{array}\right)
\hspace{1cm} \vec r = \left( \begin{array}{c}
r \sqrt{1-x^{2}} \cr 0 \cr r x
\end{array}\right)  \ \ ,
\end{equation}
where $x$ is the cosine of the angle between $\vec{r}$ and $\vec{R}$.
In addition, one should note that $\int d\Omega_{r} d\Omega_{R}=8\pi^2 \int dx$. 
Then, Eq.~(\ref{eq:ucouple}) becomes

\begin{align}
U_{\tilde{\alpha},\tilde{\alpha}' } (R) & = \sum_{LS} 2\pi (2S+1) [(2j'_{bx}+1)(2j_{bx}+1)(2j'_A+1)(2j_A+1)(2\lambda_A+1) (2\lambda_A'+1 )]^{1/2}
\left\{ \begin{array}{ccc}   l'_{a}  & \lambda'_{A} & L  \cr
s_{bx} & s_{A}  & S \cr
j'_{bx} & j'_A & J
\end{array}\right\} \nonumber \\
& \times \left\{ \begin{array}{ccc}   l_{a}  & \lambda_{A} & L  \cr
s_{bx} & s_{A}  & S \cr
j_{bx} & j_A & J
\end{array}\right\} \ \int r^2 dr \langle r \beta' | \phi_{bx}^{n'}  \rangle \langle \phi_{bx}^{n}  |  r \beta   \rangle \sum_{M_L}\langle l_a' M_L \lambda_A' 0 | L M_L \rangle \langle l_a M_L \lambda_A 0 | L M_L \rangle \\ \nonumber
& \times \int dx   Y_{l_a'}^{M_L} (\hat{r})  Y_{l_a}^{M_L*} (\hat{r}) U(\vec{r}, \vec{R}).
\end{align}

 In the particular case in which the intrinsic spins of the particles can be ignored, the equation above takes a particularly simple form
\begin{align}
U_{\tilde{\alpha},\tilde{\alpha}' } (R) & = \frac{2\pi}{2J+1}[(2\lambda_A+1)(2\lambda_A'+1)]^{1/2} \ \int r^2 dr \langle r \beta' | \phi_{bx}^{n'}  \rangle \langle \phi_{bx}^{n}  |  r \beta   \rangle \sum_{M_L}\langle l_a' M_L \lambda_A' 0 | L M_L \rangle \langle l_a M_L \lambda_A 0 | L M_L \rangle \\ \nonumber
& \times \int dx   Y_{l_a'}^{M_L} (\hat{r})  Y_{l_a}^{M_L*} (\hat{r}) U(\vec{r}, \vec{R}).
\end{align}

To solve the coupled equations of Eq.~(\ref{eq:CDCC}), one can use the technique discussed, for example, in Refs.~\cite{BAYLIS19827,DRUET201088}
with the following outgoing boundary conditions 
\begin{align}
\langle \phi^n_{bx} R\alpha| \Psi^{j_\mathrm{in}(+)} \rangle  \underset{R\to \infty}{\rightarrow} & \frac{4\pi}{k_{\alpha_0} R} e^{i\sigma_{\lambda_0}} \sum_{\tilde{\alpha}_0}\sum_{m_{\lambda_0}m_{j_A}}\langle \lambda_0 m_{\lambda_0} s_A m_{A} | j_A m_{j_A} \rangle \langle j_a m_{a} j_A m_{j_A} | J M_J \rangle i^{\lambda_0} Y_{\lambda_0}^{m_{\lambda_0}*}(\hat{k}_a) \nonumber \\
& \times \frac{i}{2} \Big[H^{(-)}_{\lambda}\delta_{\tilde{\alpha}\tilde{\alpha}_0}-\sqrt{\frac{v_\alpha}{v_{\alpha_0}}}S_{\tilde{\alpha}\tilde{\alpha}_0}H^{(+)}_{\lambda}\Big].
\end{align}
One can note that the boundary condition also depends on the incoming channel index $\tilde{\alpha}_0$, 
%where $\tilde{\alpha}_0$ stands for 
corresponding to the $\tilde{\alpha}$ index in which projectile and target are in their ground state. In practical calculations, it is convenient to separate the angular and radial parts of $\langle \phi^n_{bx} R\alpha| \Psi^{j_\mathrm{in}(+)} \rangle$, i.e., 
\begin{align}
\langle \phi^n_{bx} R\alpha| \Psi^{j_\mathrm{in}(+)} \rangle&=\sum_{\tilde{\alpha}_0}\langle \phi^n_{bx} R\alpha| \Psi^{j_\mathrm{in}\tilde{\alpha}_0(+)} \rangle \nonumber\\
&=   \frac{4\pi}{k_{\alpha_0} R} \sum_{\tilde{\alpha}_0}e^{i\sigma_{\lambda_0}} \sum_{m_{\lambda_0}m_{j_A}}\langle \lambda_0 m_{\lambda_0} s_A m_{A} | j_A m_{j_A} \rangle \langle j_a m_{a} j_A m_{j_A} | J M_J \rangle i^{\lambda_0} Y_{\lambda_0}^{m_{\lambda_0}*}(\hat{k}_a) u_{\tilde{\alpha} \tilde{\alpha}_0}(R), 
\end{align}
where the radial functions $u_{\tilde{\alpha} \tilde{\alpha}_0}(R)$ verify the coupled equations 
\begin{equation}
    \Big[ -\frac{\hbar^2}{2\mu_{aA}} \frac{d^2}{dR^2} + \frac{\hbar^2 \lambda(\lambda+1)}{2\mu_{aA}R^2} + \varepsilon_n -E \Big] u_{\tilde{\alpha} \tilde{\alpha}_0}(R)+ \sum_{\tilde{\alpha}'} U_{\tilde{\alpha}, \tilde{\alpha}'}(R) u_{\tilde{\alpha}' \tilde{\alpha}_0}(R)=0, 
\end{equation}
subject to the boundary condition 
\begin{align}
u_{\tilde{\alpha} \tilde{\alpha}_0}(R)  \underset{R\to \infty}{\rightarrow} 
\frac{i}{2} \Big[H^{(-)}_{\lambda}\delta_{\tilde{\alpha}\tilde{\alpha}_0}-\sqrt{\frac{v_\alpha}{v_{\alpha_0}}}S_{\tilde{\alpha}\tilde{\alpha}_0}H^{(+)}_{\lambda}\Big].
\end{align}

Finally, the CDCC wave function can be represented in angular momentum basis, $| rR \alpha \rangle$,
as 
\begin{align}
\langle rR \alpha | \Psi^{j_\mathrm{in}\tilde{\alpha}_0(+)} \rangle & = 
\sum_{\alpha'n } \int R'^2 dR' \langle rR\alpha | \phi_{bx}^n R' \alpha' \rangle 
\langle \phi_{bx}^n R' \alpha' | \Psi^{j_\mathrm{in}\tilde{\alpha}_0(+)} \rangle \nonumber \\
&= \sum_{n} \langle r \alpha | \phi_{bx}^n  \rangle  
\langle \phi_{bx}^n R \alpha | \Psi^{j_\mathrm{in}\tilde{\alpha}_0(+)} \rangle
\end{align}
%-------------------------------------
\subsection{The IAV model of inclusive breakup}
%-------------------------------------
\begin{figure}[tb]
\begin{center}
 {\centering \resizebox*{0.85\columnwidth}{!}{\includegraphics{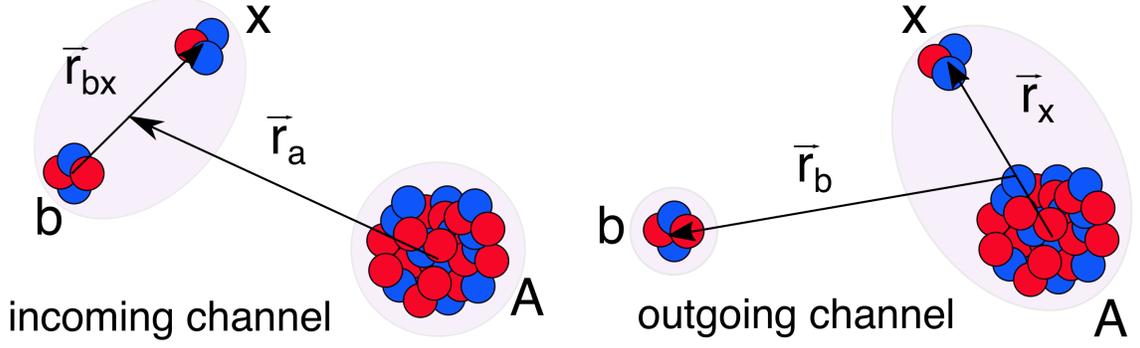}} \par}
\caption{\label{fig:fig1} Coordinates used in the breakup reaction}
\end{center}
\end{figure}
First, one can write the process under study in the form 
\begin{equation}
a(=b+x) +A \to b + B^*, 
\end{equation}
where the projectile $a$, made up of $b$ and $x$, interacts with the target $A$, leaving the particle $b$ and other fragments. Thus $B^*$ is any possible state between $x+A$ system. The relevant coordinates of these processes are depicted in Fig.~\ref{fig:fig1}. 

In the The IAV model, the inclusive breakup cross section is expressed as 
\begin{align}
\label{eq:iav_inclusive}
\frac{d^2\sigma}{dE_bd\Omega_b}=-\frac{2}{\hbar v_a} \rho_b (E_b) 
\Im \langle \rho(\vec{k}_b) |G_x^{(+)}| \rho(\vec{k}_b)\rangle ,
\end{align}
where $\rho_b(E_b)=\mu_b k_b /(8\pi^3 \hbar^2)$ is a density of states and the source term takes the form 
\begin{align}
\langle \vec{r}_x| \rho(\vec{k}_b) \rangle = \langle \vec{r}_x \chi_b^{(-)} (\vec{k}_b)| V_\mathrm{post} | \Psi^{3b(+)} \rangle, 
\end{align}
where $\chi_b^{(-)*} (\vec{k}_b,\vec{r}_{bB})$ is the distorted wave (obtained with some optical potential $U_{bB}$) describing
the relative motion between $b$ and $B$, $V_\mathrm{post}=V_{bx} + U_{bA}-U_{bB}$ 
is the post-form transition operator and $\Psi^{3b(+)}$ the three-body scattering wave function.
The imaginary part of the Green's function can be expressed as \cite{Kas82}
\begin{align}
\Im G_x^{(+)} = (1+U_x^\dagger G_x^{(+)\dagger}) \Im G_0 (1 + G_x^{(+)}U_x) + G_x^{(+)\dagger}W_x G_x^{(+)},
\end{align}
where $W_x$ is the imaginary part of optical potential $U_x$. 
Inserting the above equation into Eq.~(\ref{eq:iav_inclusive}), one gets 
\begin{align}
\label{eq:iav_inclusive2}
\frac{d^2\sigma}{dE_b d\Omega_b}&=-\frac{2}{\hbar v_a} \rho_b(E_b) 
\langle \rho(\vec{k}_b)| (1+U_x^\dagger G_x^{(+)\dagger}) \Im G_0 (1 + G_x^{(+)}U_x) | \rho(\vec{k}_b) \rangle 
-\frac{2}{\hbar v_a} \rho_b(E_b) \langle \rho(\vec{k}_b)| G_x^{(+)\dagger}W_x G_x^{(+)} | \rho(\vec{k}_b) \rangle .
\end{align}
The first and second terms correspond, respectively, the elastic and nonelastic breakup contributions of the inclusive breakup cross section. The former can be written in a more familiar form using the energy-conserving delta function operator
 %By using the energy-conserving delta function operator 
\begin{align}
\Im G_0 = -\pi \delta (E_x - \frac{k_x^2}{2\mu_{xA}}),
\end{align}
%then the first part of Eq.~(\ref{eq:iav_inclusive2})  which is also the elastic breakup part takes the form 
so that the EBU terms results
\begin{align}
\frac{d^2\sigma}{dE_b d\Omega_b}\Big|_\mathrm{EBU}& = \frac{2\pi}{\hbar v_a} \rho_b(E_b) 
\langle \rho(\vec{k}_b)| (1+U_x^\dagger G_x^{(+)\dagger}) 
\delta(E_x - \frac{k_x^2}{2\mu_{xA}}) (1 + G_x^{(+)}U_x) | \rho(\vec{k}_b) \rangle \nonumber \\
& = \frac{2\pi}{\hbar v_a} \rho_b(E_b) \int \langle \rho(\vec{k}_b)| (1+U_x^\dagger G_x^{(+)\dagger})  
| \vec{k}_x \rangle \langle \vec{k}_x | \delta(E_x - \frac{k_x^2}{2\mu_{xA}}) | \vec{k}_x'\rangle 
\langle \vec{k}_x' | (1 + G_x^{(+)}U_x) | \rho(\vec{k}_b) \rangle  (2\pi)^{-6} d\vec{k}_x d\vec{k}_x'.
\end{align}
By noting that $\langle \chi^{(-)}_{x} (\vec{k}_x')| = \langle \vec{k}'_x | (1 + G_x^{(+)}U_x) $, then 
\begin{align}
\label{eq:iav_ebu}
\frac{d^2\sigma}{dE_b d\Omega_b}\Big|_\mathrm{EBU}& = \frac{2\pi}{\hbar v_a} \rho_b(E_b)  
\int \bigg| \langle \chi^{(-)}_{x} (\vec{k}_x) | \rho(\vec{k}_b)  \rangle \bigg|^2
\delta (E_x - \frac{k^2}{2\mu_{xA}} ) (2\pi)^{-3} d\vec{k}_x \nonumber \\
& = \frac{2\pi}{\hbar v_a} \rho_b(E_b)   \rho_x(E_x)   \int \bigg| \langle \chi^{(-)}_{x} (\vec{k}_x) | \rho(\vec{k}_b)  \rangle \bigg|^2
d\Omega_{k_x} \, .
\end{align}
The nonelastic breakup differential cross section (second part of Eq.~(\ref{eq:iav_inclusive2})) takes the form 
\begin{align}
\label{eq:iav_neb}
\frac{d^2\sigma}{dE_b d\Omega_b}\Big|_\mathrm{NEB}= -\frac{2}{\hbar v_a} \rho_b(E_b) 
\langle G_x^{(+)}\rho(\vec{k_b}) |W_x|  G_x^{(+)}\rho(\vec{k_b}) \rangle . 
\end{align}
The common part of Eq.~(\ref{eq:iav_ebu}) and Eq.~(\ref{eq:iav_neb}) is the source term. 
In the next section, the partial wave expansion of the source term will be discussed. 
%-------------------------------------
\subsection{Source term expression in angular momentum basis}
%-------------------------------------
%\amm{[I think this section would be more appropriate as an appendix.]} \jl{[I prefer to leave it here.]}

The source term can be expressed in the angular momentum basis $|\gamma\rangle = | l_x(s_xs_A)s_{xA};j_{xA} m_{j_{xA}} \rangle $ index with selected spin projection of the incoming and outgoing particles as 
\begin{align}
 \langle r_x \gamma | \rho^{m_b, j_\mathrm{in}}(\vec{k}_b) \rangle = \langle r_x \gamma \chi_b^{m_b(-)}(\vec{k}_b) |V_\mathrm{post}|\Psi^{j_\mathrm{in}(+)}\rangle.
\end{align}
%\amm{Pleae, check superscript of $\chi_b$. Should it be $m_b$ instead of $mb$?}\jl{[yes, you are right, fixed]}
Here we use the CDCC wave function to approximate the three-body scattering wave function. By inserting the complete basis, one  gets 
\begin{align}
\langle r_x \gamma | \rho^{mb,j_\mathrm{in}} (\vec{k}_b)  \rangle = &\sum_{\alpha_\mathrm{out}} \int dr_b r_b^2 \langle r_x \gamma \chi_b^{m_b(-)} (\vec{k}_b)| r_x r_b \alpha_\mathrm{out} \rangle \sum_{\alpha_\mathrm{cdcc}}\int_{-1}^{1} dx V_\mathrm{post} (r_x,r_b,x,\alpha_\mathrm{cdcc})\mathcal{G}^{\mathrm{out}\gets \mathrm{in}}_{\alpha_\mathrm{cdcc},\alpha_\mathrm{out}}(r_xr_bx)  \nonumber \\ 
& \times \langle r_{bx}r_a \alpha_\mathrm{cdcc} | \Psi^{j_\mathrm{in}(+)}  \rangle,
\end{align}
with 
%\amm{$mb$ should be $m_b$?}\jl{[fixed]}
\begin{align}
\langle r_x \gamma \chi_b^{m_b(-)} (\vec{k}_b)| r_x r_b \alpha_\mathrm{out} \rangle =& \frac{4\pi}{k_b r_b} \sum_{m_{\lambda_b}}\sum_{m_{j_b}} \langle \lambda_b m_{\lambda_b} s_{b} -m_{b} | j_b m_{j_b} \rangle (-)^{s_b+m_b} \langle j_{xA} m_{j_{xA}} j_b m_{j_b}|J M \rangle \nonumber \\
& \times f_{\lambda_b}^{j_b} (k_b,r_b) e^{i\sigma_{\lambda_b}} i^{-\lambda_b} Y_{\lambda_b}^{m_{\lambda_b}} (\hat{k}_b), 
\end{align}
where $f_{\lambda_b}^{j_b} (k_b,r_b)$ is the radial part of the solution of the Schr\"odinger equation with the optical potential $U_{bB}$ and 
\begin{align}
\label{eq:galphaalpha}
\mathcal{G}^{\mathrm{out}\gets \mathrm{in}}_{\alpha_\mathrm{cdcc},\alpha_\mathrm{out}}(r_xr_bx) =& \sum_{LS} (2S+1) \sqrt{(2j_{bx}+1)(2j_{A}+1)(2j_{xA}+1)(2j_{b}+1)}
\left\{ \begin{array}{ccc}   l_{x}  & s_{xA} & j_{xA}  \cr
\lambda_{b} & s_{b}  & j_{b} \cr
L & S & J
\end{array}\right\} \,
\left\{   \begin{array}{ccc}   l_{a}  & s_{bx} & j_{bx}  \cr
\lambda_{A} & s_{A}  & j_{A} \cr
L & S & J
\end{array}\right\}
 \nonumber \\
 & \times {8\pi^{2}} \, \sum_{M_L=-L}^{L}
\left\{ Y_{l_{x}}^{m_{l_{x}}*}(\hat r_{x}) \, Y^{m_{\lambda_{b}}*}_{\lambda_{b}}(\hat r_{b})   \right\}^{LM_L}
\left\{ Y^{m_{l_{a}}}_{l_{a}}(\reallywidehat {a \vec r_{x} - \vec r_{b} })
 \, Y^{m_{\lambda_{A}}}_{\lambda_{A}}( \reallywidehat {b\vec r_{x} + c \vec r_{b} } )   \right\}^{LM_L}
\nonumber \\
&  \times  (-)^{s_{bx}+2s_{A}+s_{x}+s_{b}} \sqrt{(2s_{xA}+1)(2s_{bx}+1)}
\left\{  \begin{array}{ccc}   s_{A}  & s_{x} & s_{xA}  \cr
s_{b} & S  & s_{bx} \cr
\end{array}\right\},
\end{align}
where the angular momentum states are defined as 
\begin{eqnarray}
     | \alpha_\mathrm{cdcc}\rangle &=& | (l_a (s_b s_x)s_{bx})j_{bx} (\lambda_A s_A)j_A; J  M \rangle  \nonumber \\
     | \alpha_\mathrm{out}\rangle &=& | (l_x (s_x s_A)s_{xA})j_{xA} (\lambda_b s_b)j_b; J  M \rangle, 
\end{eqnarray}
where $l$ is the angular momentum between the  $b-x$ pair and $\lambda$ is the one between the pair and the third fragment. The coefficients $a$, $b$ and $c$ are the mass rations 
\begin{eqnarray}
a & =& \frac{m_{A}}{m_{A}+m_{x}}  \nonumber \\
b  & = & \frac{(m_{b}+m_{x}+m_{A})\, m_{x}}{(m_{A}+m_{x})(m_{b}+m_{x})} \\ \nonumber
c & =& \frac{m_{b}}{m_{b}+m_{x}} .
\end{eqnarray}
The relevant coordinates of the incoming channel are given by
\begin{eqnarray}
r_{bx}(r_{x}r_{b}x) & =& \sqrt{a^{2} r_{x}^{2} +  r_{b}^{2} - 2 a  r_{x}r_{b}x}  \\ \nonumber
r_{a}(r_{x}r_{b}x) & =& \sqrt{b^2r_{x}^{2} + c^{2} r_{b}^{2} + 2 bc r_{x}r_{b}x}  \ \ .
\end{eqnarray}

In Eq.~(\ref{eq:galphaalpha}) the curly brackets grouping the spherical harmonics indicate that they are coupled to a state of total orbital angular momentum $L$ and third component $M_L$. For the evaluation, one can choose $\hat{r}_b$ as $z$ direction and assume that $\vec{r}_x$ is in the $x-y$ plane, i.e., 
\begin{equation}
\vec r_{b} = \left( \begin{array}{c}
0 \cr 0 \cr r_{b}
\end{array}\right)
\hspace{1cm} ; \hspace{1cm} \vec r_{x} = \left( \begin{array}{c}
r_{x} \sqrt{1-x^{2}} \cr 0 \cr r_{x} x
\end{array}\right)  \ \ ,
\end{equation}
where $x$ is the cosine of the angle between $\vec{r}_b$ and $\vec{r}_x$. Then $\hat{r}_a$ and $\hat{r}_{bx}$ can be computed accordingly.

%-------------------------------------
\subsection{Elastic Breakup}
%-------------------------------------
The elastic breakup cross section Eq.~(\ref{eq:iav_ebu}) in the angular momentum basis can be rewritten as 
\begin{align}
\frac{d^2 \sigma}{dE_b d\Omega_b} \Big |_\mathrm{EBU} = \frac{2\pi}{\hbar v_a} \rho_b(E_b) \rho_x(E_x) \frac{1}{2j_a+1}\frac{1}{2s_A+1}\sum_{m_a}\sum_{m_A}\sum_{m_b}\sum_{m_x}\int \Big|\langle  \chi_x^{(-)m_x, m_A} (\vec{k}_x) | \rho^{m_b,j_\mathrm{in}}(\vec{k}_b) \rangle \Big|^2 d\Omega_x.
\end{align}

By inserting the complete basis, one obtains
\begin{align}
\frac{d^2 \sigma}{dE_b d\Omega_b} \Big |_\mathrm{EBU} = &\frac{2\pi}{\hbar v_a} \rho_b(E_b) \rho_x(E_x) \frac{1}{2j_a+1}\frac{1}{2s_A+1}
%\sum_{m_a}\sum_{m_A}\sum_{m_b}\sum_{m_x} 
\nonumber \\
& \times \sum_{m_a}\sum_{m_A}\sum_{m_b}\sum_{m_x}  \int \Big|\sum_{\gamma} \int\langle  \chi_x^{(-)m_x, m_A} (\vec{k}_x) | r_x\gamma \rangle \langle r_x\gamma| \rho^{m_b,j_\mathrm{in}}(\vec{k}_b) \rangle r_x^2 dr_x \Big|^2 d\Omega_x, 
\end{align}
where 
\begin{align}
\langle  \chi_x^{(-)m_x, m_A} (\vec{k}_x) | r_x\gamma \rangle = \frac{4\pi}{k_x r_x} f_{l_x}^{j_{xA}}(k_xr_x) e^{i\sigma_{l_x}} (-)^{s_{xA}+m_x+m_A}\sum_{m_{l_x}}\langle l_x m_{l_x} s_{xA} -(m_x+m_A)|j_{xA}m_{j_{xA}} \rangle i^{-l_x}Y_{l_x}^{m_{l_x}}(\hat{k}_x),
\end{align}
in which $f_{l_x}^{j_{xA}}(k_xr_x)$ is the radial part of the distorted wave solution for the optical potential $U_x$.

%------------------------------
\subsection{Nonelastic Breakup}
%------------------------------
In the angular momentum basis, the NEB cross section of Eq.~(\ref{eq:iav_neb}) can be rewritten as 
\begin{align}
\frac{d^2\sigma}{dE_b d\Omega_b} \Big |_\mathrm{NEB} = -\frac{2}{\hbar v_a} \rho_b(E_b) \frac{1}{2j_a+1} \frac{1}{2s_A+1}\sum_{m_a} \sum_{m_A} \sum_{m_b}\sum_{\gamma} \int r_x^2 \, |\psi^{\gamma,m_b,j_\mathrm{in}}_x (r_x,\vec{k}_b) |^2
%\Big| \langle r_x \gamma | G_x^{(+)} \rho^{m_b,j_\mathrm{in}}(\vec{k}_b) \rangle \Big|^2
W_x^{\gamma}(r_x)  dr_x , 
\end{align}
where 
\begin{align}
\label{eq:psi_xA}
\psi^{\gamma,m_b,j_\mathrm{in}}_x (r_x,\vec{k}_b) \equiv \langle r_x \gamma | G_x^{(+)} \rho^{m_b,j_\mathrm{in}}(\vec{k}_b) \rangle = \int r_x'^2 G_x^{(+)}(r_x,r_x',\gamma) \langle r_x'\gamma | \rho^{m_b,j_\mathrm{in}}(\vec{k}_b) \rangle dr_x',
\end{align}
with 
\begin{equation}
 G_x^{(+)}(r_x,r_x',\gamma) = -\frac{2\mu_x}{\hbar^2}\frac{1}{k_x r_x r_x'} f_\gamma(r_{x<})h_\gamma^{(+)} (r_{x>}), 
\end{equation}
where $r_{x<}$ means the lesser of $r_x$ and $r_x'$, $f_\gamma$ and $h_\gamma^{(+)}$ are the regular and irregular of the solution with optical potential $U_x^\gamma$, respectively. 

%\amm{What is the difference between the $R$ functiona and the $\psi$ function of eq. (46)?}\jl{Eq(46) contains CG coefficients and $Y_l^m (\hat{k})$}

To assist the discussion in the next section, we introduce also for completeness the expression for the angle-integrated NEB differential cross section, which is given by:
\begin{align}
\label{eq:dsdeb}
\frac{d\sigma}{dE_b}=\int \frac{d^2\sigma}{dE_b d\Omega_b} d\Omega_b  = -\frac{1}{2\pi\hbar v_a} \rho_b(E_b) \frac{1}{2j_a+1} \frac{1}{2s_A+1}\sum_{\alpha_\mathrm{in}}\sum_{\alpha_\mathrm{out}} \int r_x^2 \, |R^{\alpha_\mathrm{in},\alpha_\mathrm{out}}_x (r_x) |^2
%\Big| \langle r_x \gamma | G_x^{(+)} \rho^{m_b,j_\mathrm{in}}(\vec{k}_b) \rangle \Big|^2
W_x^{\gamma}(r_x)  dr_x , 
\end{align}
where
\begin{align}
\label{eq:Rx}
R^{\alpha_\mathrm{in},\alpha_\mathrm{out}}_x (r_x) 
&= \frac{16\pi^2}{k_ak_b }\int r_x'^2 G_x^{(+)}(r_x,r_x',\gamma) dr_x'  \int dr_b r_b 
 f_{\lambda_b}^{j_b} (k_b,r_b) e^{i\sigma_{\lambda_b}} i^{-\lambda_b}   \sum_{\tilde{\alpha}_\mathrm{cdcc}}\int_{-1}^{1} dx V_\mathrm{post} (r'_x,r_b,x,\alpha_\mathrm{cdcc})  \nonumber \\ 
& \times \mathcal{G}^{\mathrm{out}\gets \mathrm{in}}_{\alpha_\mathrm{cdcc},\alpha_\mathrm{out}}(r'_xr_bx) \langle r_{bx} \alpha_\mathrm{cdcc} | \phi_{bx}^n  \rangle  
\frac{u_{\tilde{\alpha}_\mathrm{cdcc} \tilde{\alpha}_\mathrm{in}}(r_a)}{r_a} e^{i\sigma_{\lambda_\mathrm{in}}}  i^{\lambda_\mathrm{in}}  .
\end{align}
where $| \alpha_\mathrm{in}\rangle = | (l_a (s_b s_x)s_{bx})j_{a} (\lambda_\mathrm{in} s_A)j_A; J  M \rangle$ is the set of quantum numbers characterizing the spin-angular part of the incident channel. 
% \amm{Ins't $\alpha_{in}$ essentially the same as $\alpha_{cdcc}$ of eq. (38)?}\jl{no, they are not the same, $\alpha_{in}$ stands for the alpha index with only the ground state of the projectile whereas $\alpha_{cdcc}$ contains the continuum states.}

%---------------------------
\section{Numerical results}
%-----------------------------------
\label{sec:III}
In this section, we assess the validity of DWBA-NEB calculations by comparing them with CDCC-NEB results for the  breakup reaction of $^{93}$Nb$(d,pX)$ at $E_d=25.5$~MeV using a toy model in which the deuteron binding energy is varied artificially.  This reaction has already been studied in our earlier work~\cite{Jin15} using the DWBA versions of the IAV model, and good agreement with the experimental data was found.
% In addition, we investigated the accuracy of DWBA computations by comparing them to the CDCC version of the IAV model~\cite{Jin19}. \amm{ I think this latter sentence is more appropriate for the introduction}.

Whenever possible, we have employed the same potentials as in our previous calculations for this reaction. In the CDCC calculations, $\ell = 0-4$ partial waves and up to a maximum excitation energy of 20~MeV were considered for $n-p$ continuum states. Intrinsic spins are disregarded to simplify the calculations. The deuteron-target potential used for the DWBA calculations was initially adopted from Ref.~\cite{Han06}, but the potential depths were readjusted to reproduce the elastic scattering differential cross section calculated by CDCC. Table~\ref{tab:tab1} lists the fitted parameters for various binding energies of the toy deuteron model. During the fitting process, the remaining parameters were kept unchanged. The rationale behind this adjustment is to limit the variability when comparing the NEB cross-sectional differentials derived by these methods.

\begin{figure}[tb]
\begin{center}
 {\centering \resizebox*{0.85\columnwidth}{!}{\includegraphics{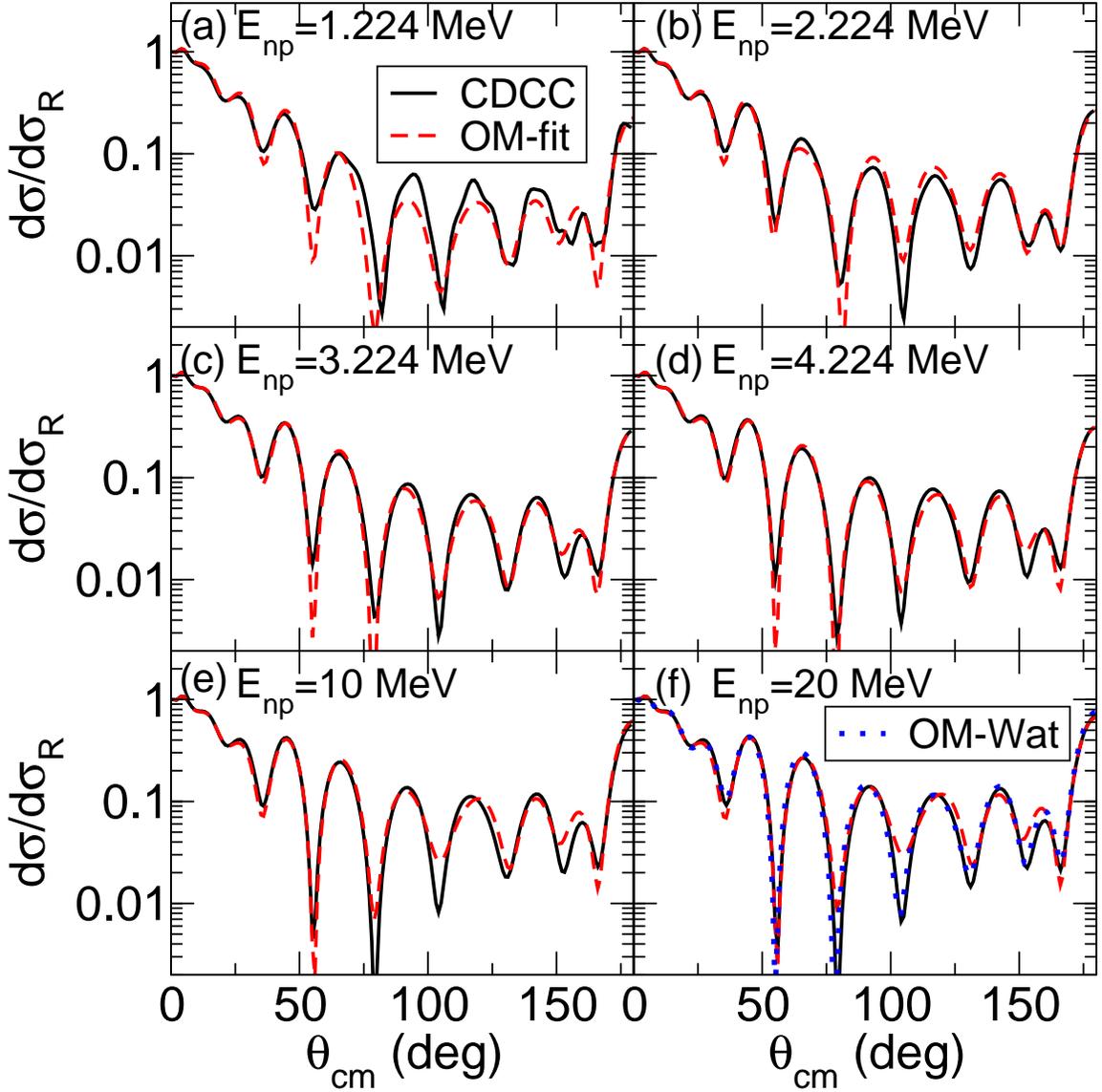}} \par}
\caption{\label{fig:fig2} Elastic scattering of $d+^{93}$Nb at 25.5 MeV with different binding energy of $n-p$ pair, $(a) E_{np}=1.224$ MeV, $(b) E_{np}=2.224$ MeV, $(c) E_{np}=3.224$ MeV,  $(d) E_{np}=4.224$ MeV, $(e) E_{np}=10$ MeV, and $(f)E_{np}=20$ MeV. }
\end{center}
\end{figure}

\begin{table}[h]
\centering
\caption{Fitted parameters for various binding energies while keeping the rest of parameters constant, namely $r_v$ = 1.17 fm, $a_v$ = 0.81 fm, $r_w$ = 1.56 fm, $a_w$ = 0.9 fm, $r_{wd}$ = 1.33 fm, and $a_{wd}$ = 0.67 fm. 
}\label{tab:tab1}
\begin{tabular}{c|c|c|c|c|c|c}
\hline
$E_{np}$ (MeV)   & $1.224$  & $2.224$ & $3.224$ & $4.224$ & $10$ & $20$  \\ 
\hline \hline
$V$ (MeV)  & 69.1                & 105.7               & 71.3                & 71.6                & 71.6             & 71.5             \\ %\hline
$W$ (MeV) & 7.9                 & 2.7                 & 3.7                 & 2.8                 & 2.3              & 1.9              \\ %\hline
$W_d$ (MeV) & 0.8                 & 11.4                & 6.2                 & 7.2                 & 6.7              & 7.1              \\ \hline
\end{tabular}
\end{table}

Figure~\ref{fig:fig2} shows the angular distribution of the elastic scattering cross section of $d+^{93}$Nb at an incident energy of 25.5 MeV for the deuteron binding energies $(a) E_{np}=1.224$ MeV, $(b)E_{np}=2.224$ MeV, $(c)E_{np}=3.224$ MeV, $(d)E_{np}=4.224$ MeV, $(e)E_{np}=10$ MeV, and $(f)E_{np}=20$ MeV. The CDCC results are depicted by solid lines, whereas the dashed lines represent the phenomenological optical model calculations with parameters adjusted to reproduce the CDCC result. Overall, we find good agreement between both calculations in the whole angular range. The OM so obtained was used to compute the entrance channel distorted waves in the DWBA-NEB calculations [c.f.~Eq.~(\ref{eq:psi_xA})]. 
It is worth noting that the effect of the deuteron continuum on the elastic cross section becomes progressively smaller as the binding energy increases, becoming negligible for the well bound cases. This is illustrated in Fig.~\ref{fig:fig2}(f) where, in addition to the full CDCC calculation, we also include (dotted line) the CDCC calculation omitting the p-n unbound states, which  is equivalent to an optical model calculation with the so-called Watanabe potential~\cite{Watanabe58}, i.e., the diagonal ground state potential from the CDCC calculation.

\begin{figure}[htb!]
\begin{center}
 {\centering \resizebox*{0.85\columnwidth}{!}{\includegraphics{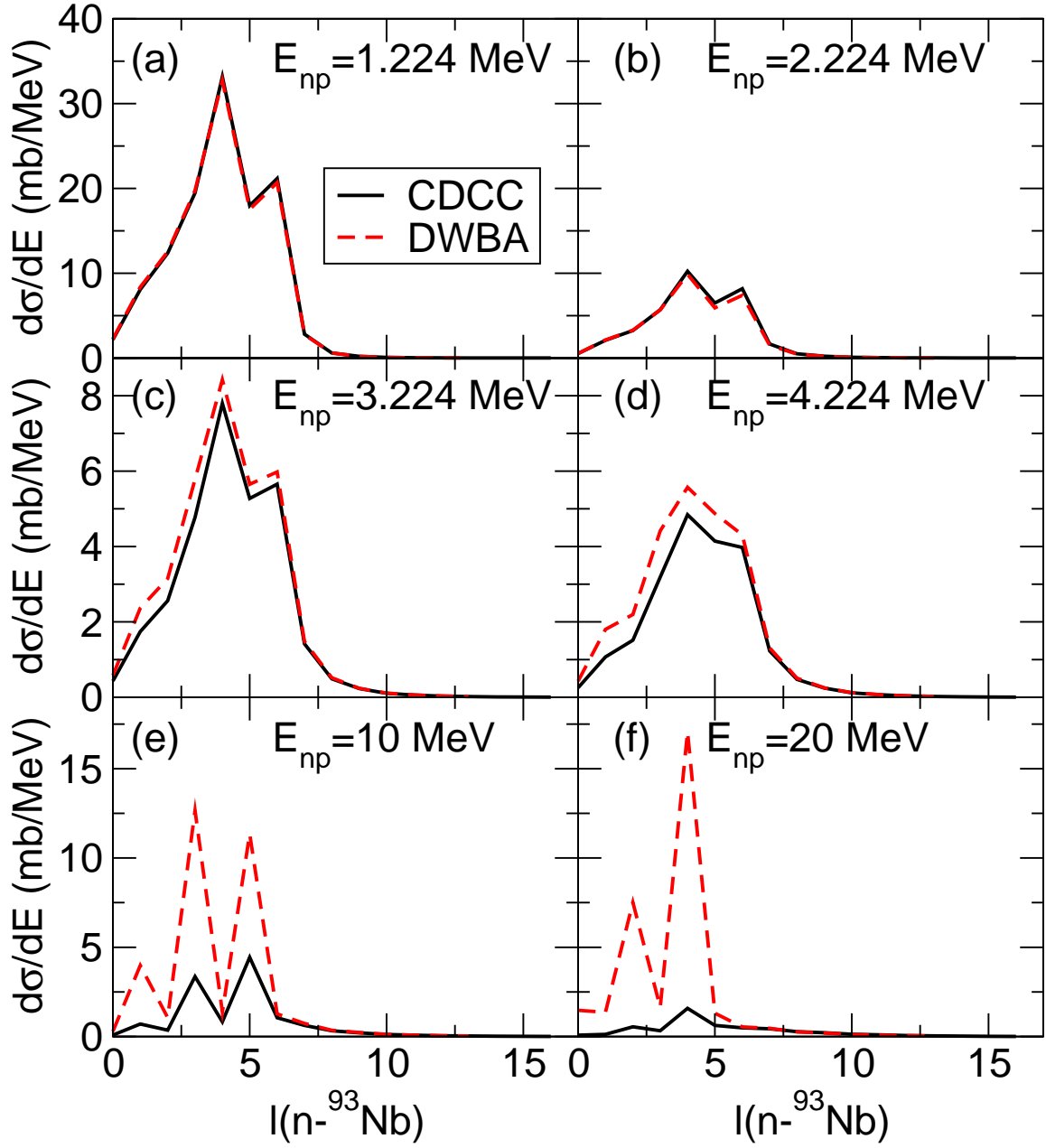}} \par}
\caption{\label{fig:fig3} The NEB contribution to the $^{93}$Nb$(d,pX)$ reaction at a laboratory energy of 25.5 MeV for an outgoing proton center-of-mass energy of 14 MeV versus the neutron-target orbital angular momentum. Each panel corresponds to an assumed binding energy of the deuteron, as indicated by the labels.}
\end{center}
\end{figure}

\begin{figure}[htb!]
\begin{center}
 {\centering \resizebox*{0.85\columnwidth}{!}{\includegraphics{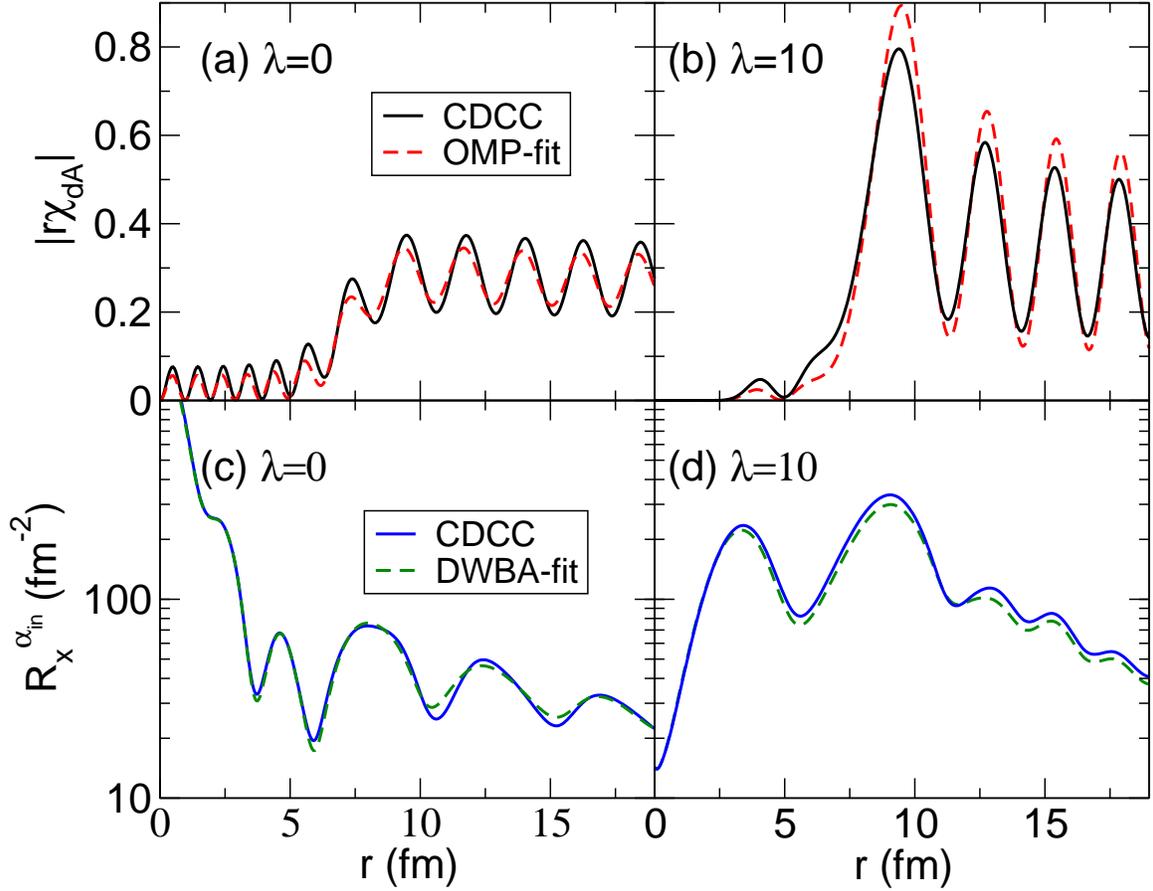}} \par}
\caption{\label{fig:fig3.5} Top panels: Modulus of  the reduced radial part of elastic channel wave function for d+$^{93}$Nb at $E=25.5$~MeV for the incoming partial waves $\lambda=0$ (a) and $\lambda=10$ (b). Bottom panels: Modulus of the radial part of neutron-target wave function defined in Eq.~(\ref{eq:psi_xA})  for $\lambda=0$ (c) and $\lambda=10$ (d).}
\end{center}
\end{figure}

\begin{figure}[tb]
\begin{center}
 {\centering \resizebox*{0.85\columnwidth}{!}{\includegraphics{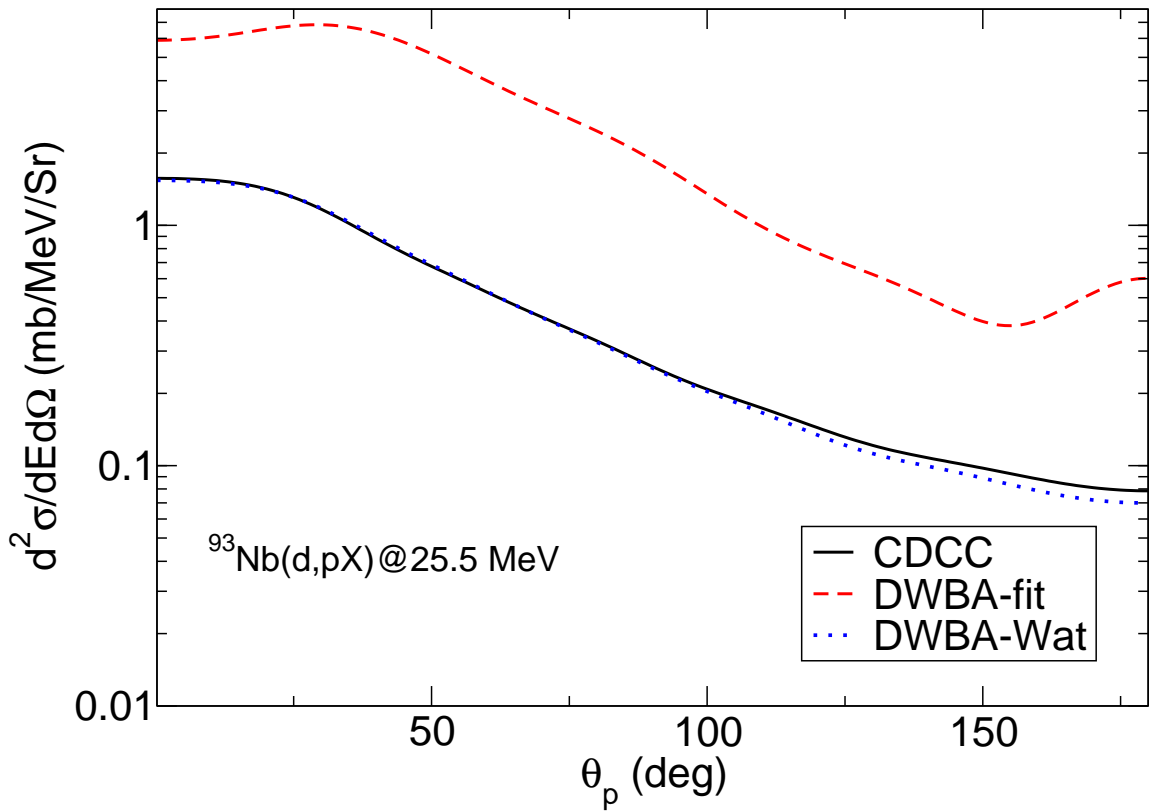}} \par}
\caption{\label{fig:fig4} NEB double differential cross section angular distribution for the $d+^{93}$Nb reaction at 25.5~MeV and an outgoing proton energy of 14 MeV in the center-of-mass frame, using a deuteron binding energy of 20 MeV. }
\end{center}
\end{figure}

In Fig.~\ref{fig:fig3}, we present the angle-integrated NEB differential cross section, $d\sigma/dE_p$, as a function of the neutron-target orbital angular momentum at a proton energy of $E_p=14$ MeV in the center-of-mass (c.m.) frame [c.f.~Eq.~(\ref{eq:dsdeb})]. The solid and dashed lines correspond, respectively, to the CDCC-NEB and  DWBA-NEB calculations. Panels (a) to (f) correspond to the deuteron separation energies 1.224 MeV, 2.224 MeV, 3.224 MeV, 4.224 MeV, 10 MeV, and 20 MeV, respectively.  Several conclusions can be drawn from this figure. First, one sees that the NEB cross section decreases as the binding energy increases. Specifically, a significant change can be observed by comparing the binding energies $E_{np}=1.224$~MeV and $E_{np}=2.224$~MeV. This is because the former represents a loosely bound halo-like case, which will be extremely vulnerable to breakup during a collision. Second,  the CDCC-NEB and DWBA-NEB calculations are found to yield almost identical results for the least bound case, but they progressively depart from each other as the binding energy grows.

To further investigate these results, we analyze in more detail the cases $E_{np}=2.224$~MeV and $E_{np}=20$~MeV, as representative examples of the weakly bound and strongly bound cases. In the top panels of Fig.~\ref{fig:fig3.5}, we present the modulus of the elastic channel wave function ($\chi_{aA}^{(+)}$) for the partial waves $\lambda=0$ (a) and $\lambda=10$ (b) and a binding energy of $E_{np}=2.224$ MeV, where $\lambda$ refers to the orbital angular momentum between the deuteron and the target. In each plot, we include the results obtained with the phenomenological optical potential (OMP-fit) and with the CDCC calculation. It is apparent that both calculations agree well in phase and magnitude, both in the asymptotic and internal regions. In the bottom panels of this figure, we show the quantity  $R_x^{\alpha_\mathrm{in}}=\frac{1}{2J+1}\sum_{\alpha_\mathrm{out}}|R_x^{\alpha_\mathrm{in},\alpha_\mathrm{out}}|^2$, 
%where $| \alpha_\mathrm{in}\rangle = | (l_a (s_b s_x)s_{bx})j_{a} (\lambda_\mathrm{in} s_A)j_A; J  M \rangle$ and 
where $R_x^{\alpha_\mathrm{in},\alpha_\mathrm{out}}$ is the  radial part of he neutron-target
wave function  given by Eq.~(\ref{eq:Rx}). 
The  quantity $R_x^{\alpha_\mathrm{in}}$ is meant to provide information on the spatial location of the NEB cross section along the $x$-A coordinate.
%\amm{[Is this $\lambda$ (d-A) or $\l_n$? In any case, I think we must specify both angular momenta, right?]}\jl{[this is only for the angular momentum between deuteron and target. The difference comes from the incoming wave function, I think for plotting the relative angular momentum for this coordinate makes more sense. ]}
Again, the solid and dashed lines represent the results obtained evaluating the source term with the  CDCC wavefunctionn or in the DWBA approximation with the phenomenological optical potential, respectively. %\amm{[In fig. 3 the solid lines were for cdcc and dashed for OMP. I think we should use a consistent convention throughout the paper]}\jl{[fixed]} 
The wavefunction calculated with both methods are remarkably close, which explains the results in Fig.~\ref{fig:fig3}(b).

% We now consider the well-bound case, $E_{np}=20$ MeV. In Fig.~\ref{fig:fig4}(a), we compare the elastic differential cross section computed with the CDCC method (solid line) and with the deuteron optical potential adjusted to the CDCC result (dashed line).   To highlight the effect of the continuum, we also include an optical model calculation performed with the Watanabe potential, i.e., the diagonal ground state potential from the CDCC calculation (dotted line). This can be regarded as a CDCC calculation in which the unbound states of the p-n system are excluded.  \amm{[This panel is essentially the same as fig. 2(f) (except for the inclusion of the Watanabe result) so I think we should remove it (if needed, we can include the Watanabe result in fig 2(f)]}
 
% we start by exploring the angular distribution of elastic scattering within the CDCC framework. as shown by the solid line in Fig.~\ref{fig:fig4}(a). \amm{ As in the previous cases, we compare calculations in which the entrance channel is described either with the CDCC wave function or with a deuteron optical potential adjusted to the CDCC elastic scattering. }\jl{Furthermore, to demonstrate the role of the continuum, we conduct an optical model calculation using Watanabe potential~\cite{Watanabe58}. The Watanabe potential calculation can be thought of as a CDCC calculation in which the unbound states of the deuteron are excluded.} 
We now consider the well-bound case, $E_{np}=20$ MeV. The elastic scattering distributions predicted by the three calculations are remarkably similar, as shown in Fig.~\ref{fig:fig2}(f). However, the NEB results, shown in Fig.~\ref{fig:fig4}, exhibit significant differences, with the DWBA-IAV NEB cross section exceeding greatly the CDCC-IAV result. To pin down the effect of the continuum states in the CDCC-IAV calculation, we also show the NEB cross section computed with the DWBA-IAV formula but with the Watanabe potential. This calculation is very close to the full CDCC-IAV result, confirming the fat that continuum states are not responsible for the disagreement between the CDCC-IAV and DWBA-fit calculations.  

These findings indicate that the description of the entrance channel wavefunction can drastically affect the calculated NEB cross sections. In the particular case analyzed here, we consider that the CDCC wavefunction, being based on the better known nucleon-nucleus potentials, should provide more reliable results. Indeed, we cannot rule out the possibility that an alternative choice of the phenomenological optical potential may give results closer to those obtained with CDCC, but we have not explored this possibility further in this work.

%suggest that the comparison of the DWBA and CDCC results significantly depends on the optical potential between the projectile and the target.

\begin{figure}[hbt!]
\begin{center}
 {\centering \resizebox*{0.85\columnwidth}{!}{\includegraphics{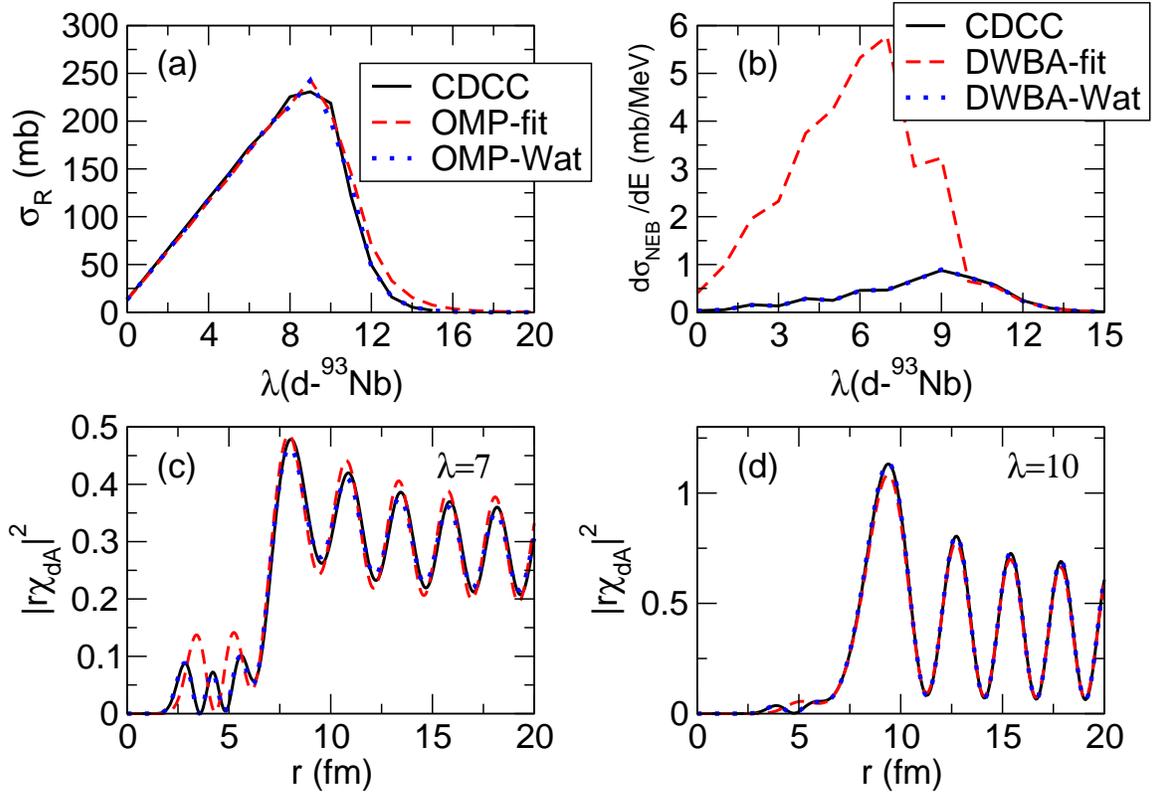}} \par}
\caption{\label{fig:fig5} (a) Partial wave distributions of reaction cross-section between projectile and target  $d+^{93}$Nb at $E=25.5$~MeV with a (hypothetical) deuteron binding energy of 20 MeV; (b) Partial wave distributions of NEB cross-section (with $\lambda$ the projectile-target relative angular momentum); (c) and (d) Modulus squared of the reduced 
radial part of the  $d+^{93}$Nb  elastic scattering wave function for $\lambda=7$ and  $\lambda=10$, respectively. }
\end{center}
\end{figure}

To clarify the differences between the DWBA-IAV and CDCC-IAV results we have performed some additional calculations, summarized in Fig.~\ref{fig:fig5}.
%\sout{Figure~\ref{fig:fig5}(a) displays the partial wave distributions of the reaction cross-section using the OM, CDCC and CDCC-gs calculations described previously.}
Figure \ref{fig:fig5}(a) presents the partial wave distributions of the reaction cross section obtained with the  CDCC approach, with the phenomenological optical model  (OMP-fit), and with the Watanabe potential (OMP-Wat) mentioned earlier. The three calculations give very similar results. This is consistent with the results shown in Figure~\ref{fig:fig4}(a) for the elastic scattering cross section. Figure~\ref{fig:fig5}(b) displays the partial wave distribution of the NEB differential cross section energy distribution, $d\sigma/dE$, for  a CM frame outgoing proton energy of 14~MeV. The CDCC and DWBA results agree  only for partial waves with $\lambda>10$, while significant differences are evident at lower partial waves. Note that the DWBA calculation utilizing the Watanabe potential produces comparable results to the full CDCC calculation. 

\begin{figure}[hbt!]
\begin{center}
 {\centering \resizebox*{0.85\columnwidth}{!}{\includegraphics{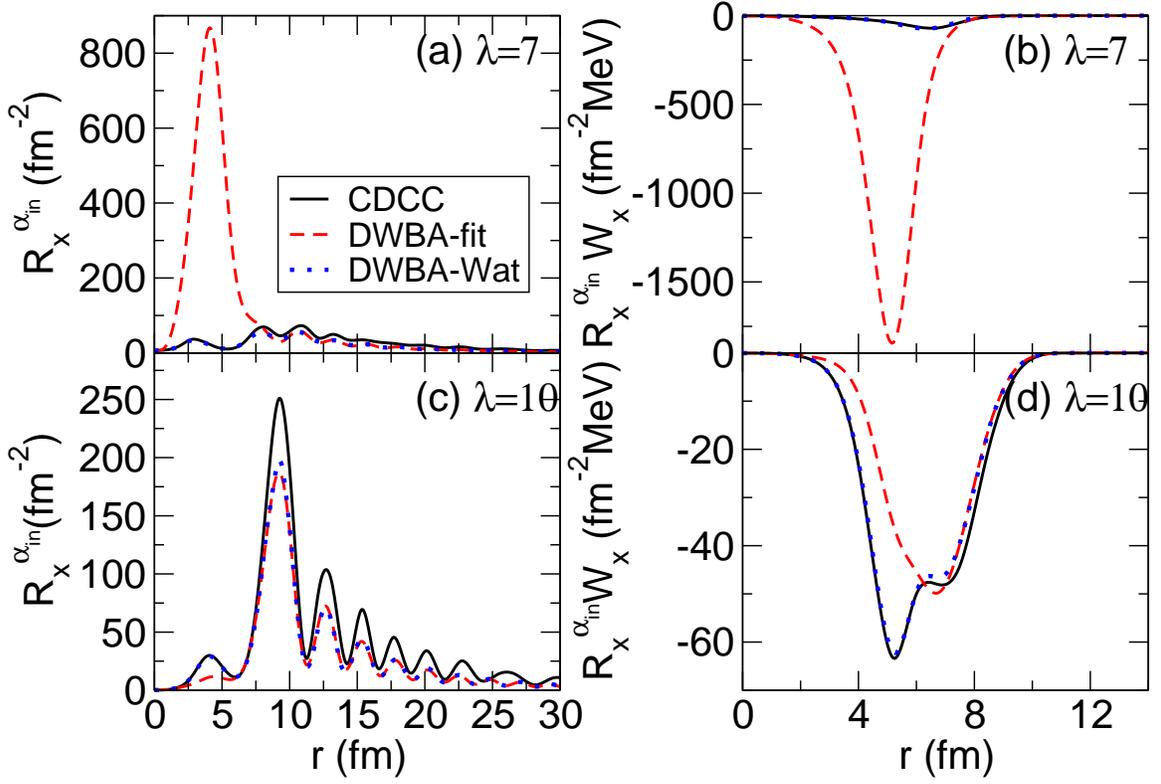}} \par}
\caption{\label{fig:fig6} Panels (a) and (c):  Radial dependence of the quantity  $R_x^{\alpha_\mathrm{in}}$ (see main text) as a function of the  neutron-target separation for the  $d+^{93}$Nb  reaction at $E=25.5$~MeV and an artificial deuteron binding energy of 20~MeV  for the projectile-target orbital angular momenta $\lambda=7$ (a) and  $\lambda=10$ (c). Panels (b) and (d): Product  $R_x^{\alpha_\mathrm{in}}(r_x) \, W_x(r_x)$, with $W_x(r_x)$ the imaginary part of the neutron-target optical potential.}
\end{center}
\end{figure}

The similarity of the $d+A$ elastic and reaction cross section computed with the OMP-fit and CDCC methods suggests that their elastic $d+^{93}$Nb wavefunctions must agree in the asymptotic region. However, these functions do not necessarily coincide in the internal region, and this might explain the differences in the NEB cross sections. To explore this, in Figs.~\ref{fig:fig5}(c) and (d), the modulus of elastic scattering wavefunctions computed by CDCC, the Watanabe OMP and the fitted OMP (OMP-fit) are shown for $\lambda=7$ and $\lambda=10$ waves, respectively. The results confirm that the asymptotic part of these  wave functions  agree (both in phase and magnitude) for all three calculations. However, the CDCC and  OMP-fit wavefunctions differ considerably in the internal region. This result suggests that the discrepancy seen in Fig.~\ref{fig:fig5}(b) is primarily due to variations in the internal part of the elastic wave function. Specifically, the difference is maximum for $\lambda=7$, while for $\lambda=10$ the interior part of the wave function is relatively small compared to the external part, resulting in a small difference in the NEB cross sections.

To investigate how the differences between the elastic deuteron-target wavefunction calculated with the different models can affect the corresponding NEB cross sections, we compare in Figs.~\ref{fig:fig6}(a) and (c) the functions $R_x^{\alpha_\mathrm{in}}(r_x)$, defined earlier, for the $\lambda=7$ and $\lambda=10$  partial waves, respectively. We also show in panels (b) and (d) the product $R_x^{\alpha_\mathrm{in}}(r_x)\, W_x(r_x)$  evaluated for the same neutron-target  partial waves. The plots show that,  for $\lambda=7$, the CDCC and Watanabe potential models produce very close results within the range of the imaginary part  of the  neutron-target optical potential, as shown in Fig.~\ref{fig:fig6}(b). By contrast, the results obtained with the phenomenological optical model potential  exhibit considerable disparity when compared to the previous results. These observations align well with the results presented in Fig.~\ref{fig:fig5}(b) for the NEB cross section, where the DWBA-fit calculation deviates significantly from the CDCC and DWBA-Wat results.  For the $\lambda=10$ partial wave, although all three calculations yield slightly different results, the variations are much smaller than for $\lambda=7$ and this explains the similarity of the corresponding NEB cross sections found in Fig.~\ref{fig:fig5}(b).

%---------------------------------
\section{summary and conclusions}
%---------------------------------
\label{sec:IV}
In this work, we have presented the numerical implementation of the IAV model for the inclusive breakup of two-body projectiles using a full three-body description of the scattering problem. We have provided a detailed derivation of the IAV model with the CDCC three-body wave function, employing an angular momentum channel basis. We have provided detailed formulas for the CDCC-IAV formulation within a  unified formalism.

The method has been then applied to the d+$^{93}$Nb reaction at an incident energy of 25.5~MeV.
Our numerical application examined the difference between the CDCC-IAV and DWBA-IAV models, by artificially modifying the binding energy of the deuteron.  For the DWBA calculations, two distinct types of optical potentials have been used: one  fitted to the elastic scattering cross section obtained from the CDCC calculations and the other computed with the Watanabe model.
%Concerning the DWBA-IAV model, we utilized two distinct types of optical potential: one that was fitted to the elastic scattering cross section obtained from the CDCC calculations and the other computed with the Watanabe model.}
Our main findings can be summarized as follows:
\begin{itemize}
\item[(i)] For the weakly-bound cases ($E_{pn} \lesssim 4$~MeV), the NEB cross sections computed with the CDCC and the phenomenological OMP give very similar NEB cross sections. %Moreover, the DWBA-IAV calculation with the Watanabe potential gives a result very close to the CDCC-IAV calculation, indicating that the deuteron continuum plays a minor role in the computed NEB cross sections. 
\item [(ii)] As the deuteron separation energy is gradually increased, the CDCC-IAV and DWBA-IAV calculations deviate from each other, in spite of the fact that they give very similar elastic cross sections. 
\item [(iii)] Inspection of the elastic channel wavefunction revealed that, for the weakly bound case, these functions are very similar for the CDCC and  phenomenological OMP, both in the internal and asymptotic parts. By contrast, for the well bound cases, the elastic wavefunctions agree only in the asymptotic region, but they differ considerably in the interior. This explains the agreement of the elastic cross sections and the disagreement on the NEB cross sections since these observables are mostly sensitive  to the asymptotic and internal parts, respectively. 
\item [(iv)] To investigate the effect of the deuteron continuum, DWBA-IAV calculations with the Watanabe potential have been also performed. The elastic and NEB cross sections obtained in this case are very close the CDCC results, indicating that the continuum plays a minor role in the NEB calculated cross sections and it is therefore not responsible for the observed differences between the CDCC-IAV and DWBA-IAV NEB calculations.
\end{itemize}

%We found that both the Watanabe potential and the fitted optical potential provide reasonable results when compared to CDCC, but display substantial differences in the interior part of the wave function, where the CDCC results align closely with Watanabe's potential; however, this is not the case with the fitted optical potential. 

Although the reliability and accuracy of the CDCC-IAV and DWBA-IAV methods should be further assessed against experimental data for different systems and energies, we may foresee that, at least in the case of deuteron induced reactions,  the CDCC wavefunction should provide a more realistic description of the entrance channel wavefunction and hence more reliable NEB predictions since this method relies on nucleon-nucleus potentials, which are in general better constrained than those for nucleus-nucleus scattering.

Whereas our results have provided significant insights, our study has limitations.  In particular, for very weakly-bound nuclei, such as halo nuclei, new effects might arise due to the enhanced breakup probability. Because of the larger model space required in this case, it becomes more demanding computationally. Calculations are nevertheless underway and the results will be published elsewhere.

%\section*{Acknowledgements}

\begin{acknowledgments}
This work has been partially supported by National Natural Science Foundation of China (Grants No.12105204), by the Fundamental Research Funds for the Central Universities, by the I+D+i project Nb.~PID2020-114687GB-I00, funded by MCIN/AEI/10.13039/501100011033 and by the grant Group FQM-160 and the  project PAIDI 2020 Nb.\ P20\_01247, funded by the Consejer\'{\i}a de Econom\'{\i}a, Conocimiento, Empresas y Universidad, Junta de Andaluc\'{\i}a (Spain) and by ``ERDF A way of making Europe''.
\end{acknowledgments}

\bibliography{inclusive_prc.bib}

\begin{thebibliography}{31}%
\makeatletter
\providecommand \@ifxundefined [1]{%
 \@ifx{#1\undefined}
}%
\providecommand \@ifnum [1]{%
 \ifnum #1\expandafter \@firstoftwo
 \else \expandafter \@secondoftwo
 \fi
}%
\providecommand \@ifx [1]{%
 \ifx #1\expandafter \@firstoftwo
 \else \expandafter \@secondoftwo
 \fi
}%
\providecommand \natexlab [1]{#1}%
\providecommand \enquote  [1]{``#1''}%
\providecommand \bibnamefont  [1]{#1}%
\providecommand \bibfnamefont [1]{#1}%
\providecommand \citenamefont [1]{#1}%
\providecommand \href@noop [0]{\@secondoftwo}%
\providecommand \href [0]{\begingroup \@sanitize@url \@href}%
\providecommand \@href[1]{\@@startlink{#1}\@@href}%
\providecommand \@@href[1]{\endgroup#1\@@endlink}%
\providecommand \@sanitize@url [0]{\catcode `\\12\catcode `\$12\catcode
  `\&12\catcode `\#12\catcode `\^12\catcode `\_12\catcode `\%12\relax}%
\providecommand \@@startlink[1]{}%
\providecommand \@@endlink[0]{}%
\providecommand \url  [0]{\begingroup\@sanitize@url \@url }%
\providecommand \@url [1]{\endgroup\@href {#1}{\urlprefix }}%
\providecommand \urlprefix  [0]{URL }%
\providecommand \Eprint [0]{\href }%
\providecommand \doibase [0]{http://dx.doi.org/}%
\providecommand \selectlanguage [0]{\@gobble}%
\providecommand \bibinfo  [0]{\@secondoftwo}%
\providecommand \bibfield  [0]{\@secondoftwo}%
\providecommand \translation [1]{[#1]}%
\providecommand \BibitemOpen [0]{}%
\providecommand \bibitemStop [0]{}%
\providecommand \bibitemNoStop [0]{.\EOS\space}%
\providecommand \EOS [0]{\spacefactor3000\relax}%
\providecommand \BibitemShut  [1]{\csname bibitem#1\endcsname}%
\let\auto@bib@innerbib\@empty
%</preamble>
\bibitem [{\citenamefont {Duan}\ \emph {et~al.}(2022)\citenamefont {Duan},
  \citenamefont {Yang}, \citenamefont {Lei}, \citenamefont {Wang},
  \citenamefont {Sun}, \citenamefont {Pang}, \citenamefont {Wang},
  \citenamefont {Liu}, \citenamefont {Xu}, \citenamefont {Ma}, \citenamefont
  {Ma}, \citenamefont {Bai}, \citenamefont {Hu}, \citenamefont {Gao},
  \citenamefont {Xu}, \citenamefont {Lin}, \citenamefont {Jia}, \citenamefont
  {Ma}, \citenamefont {Sun}, \citenamefont {Wang}, \citenamefont {Yang},
  \citenamefont {Jin}, \citenamefont {Ren}, \citenamefont {Zhang},
  \citenamefont {Zhou}, \citenamefont {Hu},\ and\ \citenamefont
  {Xu}}]{Duan2022}%
  \BibitemOpen
  \bibfield  {author} {\bibinfo {author} {\bibfnamefont {F.~F.}\ \bibnamefont
  {Duan}}, \bibinfo {author} {\bibfnamefont {Y.~Y.}\ \bibnamefont {Yang}},
  \bibinfo {author} {\bibfnamefont {J.}~\bibnamefont {Lei}}, \bibinfo {author}
  {\bibfnamefont {K.}~\bibnamefont {Wang}}, \bibinfo {author} {\bibfnamefont
  {Z.~Y.}\ \bibnamefont {Sun}}, \bibinfo {author} {\bibfnamefont {D.~Y.}\
  \bibnamefont {Pang}}, \bibinfo {author} {\bibfnamefont {J.~S.}\ \bibnamefont
  {Wang}}, \bibinfo {author} {\bibfnamefont {X.}~\bibnamefont {Liu}}, \bibinfo
  {author} {\bibfnamefont {S.~W.}\ \bibnamefont {Xu}}, \bibinfo {author}
  {\bibfnamefont {J.~B.}\ \bibnamefont {Ma}}, \bibinfo {author} {\bibfnamefont
  {P.}~\bibnamefont {Ma}}, \bibinfo {author} {\bibfnamefont {Z.}~\bibnamefont
  {Bai}}, \bibinfo {author} {\bibfnamefont {Q.}~\bibnamefont {Hu}}, \bibinfo
  {author} {\bibfnamefont {Z.~H.}\ \bibnamefont {Gao}}, \bibinfo {author}
  {\bibfnamefont {X.~X.}\ \bibnamefont {Xu}}, \bibinfo {author} {\bibfnamefont
  {C.~J.}\ \bibnamefont {Lin}}, \bibinfo {author} {\bibfnamefont {H.~M.}\
  \bibnamefont {Jia}}, \bibinfo {author} {\bibfnamefont {N.~R.}\ \bibnamefont
  {Ma}}, \bibinfo {author} {\bibfnamefont {L.~J.}\ \bibnamefont {Sun}},
  \bibinfo {author} {\bibfnamefont {D.~X.}\ \bibnamefont {Wang}}, \bibinfo
  {author} {\bibfnamefont {G.}~\bibnamefont {Yang}}, \bibinfo {author}
  {\bibfnamefont {S.~Y.}\ \bibnamefont {Jin}}, \bibinfo {author} {\bibfnamefont
  {Z.~Z.}\ \bibnamefont {Ren}}, \bibinfo {author} {\bibfnamefont {Y.~H.}\
  \bibnamefont {Zhang}}, \bibinfo {author} {\bibfnamefont {X.~H.}\ \bibnamefont
  {Zhou}}, \bibinfo {author} {\bibfnamefont {Z.~G.}\ \bibnamefont {Hu}}, \ and\
  \bibinfo {author} {\bibfnamefont {H.~S.}\ \bibnamefont {Xu}} (\bibinfo
  {collaboration} {RIBLL Collaboration}),\ }\href {\doibase
  10.1103/PhysRevC.105.034602} {\bibfield  {journal} {\bibinfo  {journal}
  {Phys. Rev. C}\ }\textbf {\bibinfo {volume} {105}},\ \bibinfo {pages}
  {034602} (\bibinfo {year} {2022})}\BibitemShut {NoStop}%
\bibitem [{\citenamefont {{Di Pietro}}\ \emph {et~al.}(2019)\citenamefont {{Di
  Pietro}}, \citenamefont {Moro}, \citenamefont {Lei},\ and\ \citenamefont {{de
  Diego}}}]{Pietro19}%
  \BibitemOpen
  \bibfield  {author} {\bibinfo {author} {\bibfnamefont {A.}~\bibnamefont {{Di
  Pietro}}}, \bibinfo {author} {\bibfnamefont {A.}~\bibnamefont {Moro}},
  \bibinfo {author} {\bibfnamefont {J.}~\bibnamefont {Lei}}, \ and\ \bibinfo
  {author} {\bibfnamefont {R.}~\bibnamefont {{de Diego}}},\ }\href {\doibase
  https://doi.org/10.1016/j.physletb.2019.134954} {\bibfield  {journal}
  {\bibinfo  {journal} {Physics Letters B}\ }\textbf {\bibinfo {volume}
  {798}},\ \bibinfo {pages} {134954} (\bibinfo {year} {2019})}\BibitemShut
  {NoStop}%
\bibitem [{\citenamefont {Pesudo}\ \emph {et~al.}(2017)\citenamefont {Pesudo}
  \emph {et~al.}}]{Pes17}%
  \BibitemOpen
  \bibfield  {author} {\bibinfo {author} {\bibfnamefont {V.}~\bibnamefont
  {Pesudo}} \emph {et~al.},\ }\href {\doibase 10.1103/PhysRevLett.118.152502}
  {\bibfield  {journal} {\bibinfo  {journal} {Phys. Rev. Lett.}\ }\textbf
  {\bibinfo {volume} {118}},\ \bibinfo {pages} {152502} (\bibinfo {year}
  {2017})}\BibitemShut {NoStop}%
\bibitem [{\citenamefont {Duan}\ \emph {et~al.}(2020)\citenamefont {Duan},
  \citenamefont {Yang}, \citenamefont {Wang}, \citenamefont {Moro},
  \citenamefont {Guimarães}, \citenamefont {Pang}, \citenamefont {Wang},
  \citenamefont {Sun}, \citenamefont {Lei}, \citenamefont {{Di Pietro}},
  \citenamefont {Liu}, \citenamefont {Yang}, \citenamefont {Ma}, \citenamefont
  {Ma}, \citenamefont {Xu}, \citenamefont {Bai}, \citenamefont {Sun},
  \citenamefont {Hu}, \citenamefont {Lou}, \citenamefont {Xu}, \citenamefont
  {Li}, \citenamefont {Jin}, \citenamefont {Ong}, \citenamefont {Liu},
  \citenamefont {Yao}, \citenamefont {Qi}, \citenamefont {Lin}, \citenamefont
  {Jia}, \citenamefont {Ma}, \citenamefont {Sun}, \citenamefont {Wang},
  \citenamefont {Zhang}, \citenamefont {Zhou}, \citenamefont {Hu},\ and\
  \citenamefont {Xu}}]{Duan20}%
  \BibitemOpen
  \bibfield  {author} {\bibinfo {author} {\bibfnamefont {F.}~\bibnamefont
  {Duan}}, \bibinfo {author} {\bibfnamefont {Y.}~\bibnamefont {Yang}}, \bibinfo
  {author} {\bibfnamefont {K.}~\bibnamefont {Wang}}, \bibinfo {author}
  {\bibfnamefont {A.}~\bibnamefont {Moro}}, \bibinfo {author} {\bibfnamefont
  {V.}~\bibnamefont {Guimarães}}, \bibinfo {author} {\bibfnamefont
  {D.}~\bibnamefont {Pang}}, \bibinfo {author} {\bibfnamefont {J.}~\bibnamefont
  {Wang}}, \bibinfo {author} {\bibfnamefont {Z.}~\bibnamefont {Sun}}, \bibinfo
  {author} {\bibfnamefont {J.}~\bibnamefont {Lei}}, \bibinfo {author}
  {\bibfnamefont {A.}~\bibnamefont {{Di Pietro}}}, \bibinfo {author}
  {\bibfnamefont {X.}~\bibnamefont {Liu}}, \bibinfo {author} {\bibfnamefont
  {G.}~\bibnamefont {Yang}}, \bibinfo {author} {\bibfnamefont {J.}~\bibnamefont
  {Ma}}, \bibinfo {author} {\bibfnamefont {P.}~\bibnamefont {Ma}}, \bibinfo
  {author} {\bibfnamefont {S.}~\bibnamefont {Xu}}, \bibinfo {author}
  {\bibfnamefont {Z.}~\bibnamefont {Bai}}, \bibinfo {author} {\bibfnamefont
  {X.}~\bibnamefont {Sun}}, \bibinfo {author} {\bibfnamefont {Q.}~\bibnamefont
  {Hu}}, \bibinfo {author} {\bibfnamefont {J.}~\bibnamefont {Lou}}, \bibinfo
  {author} {\bibfnamefont {X.}~\bibnamefont {Xu}}, \bibinfo {author}
  {\bibfnamefont {H.}~\bibnamefont {Li}}, \bibinfo {author} {\bibfnamefont
  {S.}~\bibnamefont {Jin}}, \bibinfo {author} {\bibfnamefont {H.}~\bibnamefont
  {Ong}}, \bibinfo {author} {\bibfnamefont {Q.}~\bibnamefont {Liu}}, \bibinfo
  {author} {\bibfnamefont {J.}~\bibnamefont {Yao}}, \bibinfo {author}
  {\bibfnamefont {H.}~\bibnamefont {Qi}}, \bibinfo {author} {\bibfnamefont
  {C.}~\bibnamefont {Lin}}, \bibinfo {author} {\bibfnamefont {H.}~\bibnamefont
  {Jia}}, \bibinfo {author} {\bibfnamefont {N.}~\bibnamefont {Ma}}, \bibinfo
  {author} {\bibfnamefont {L.}~\bibnamefont {Sun}}, \bibinfo {author}
  {\bibfnamefont {D.}~\bibnamefont {Wang}}, \bibinfo {author} {\bibfnamefont
  {Y.}~\bibnamefont {Zhang}}, \bibinfo {author} {\bibfnamefont
  {X.}~\bibnamefont {Zhou}}, \bibinfo {author} {\bibfnamefont {Z.}~\bibnamefont
  {Hu}}, \ and\ \bibinfo {author} {\bibfnamefont {H.}~\bibnamefont {Xu}},\
  }\href {\doibase https://doi.org/10.1016/j.physletb.2020.135942} {\bibfield
  {journal} {\bibinfo  {journal} {Physics Letters B}\ }\textbf {\bibinfo
  {volume} {811}},\ \bibinfo {pages} {135942} (\bibinfo {year}
  {2020})}\BibitemShut {NoStop}%
\bibitem [{\citenamefont {Yang}\ \emph {et~al.}(2018)\citenamefont {Yang},
  \citenamefont {Liu}, \citenamefont {Pang}, \citenamefont {Patel},
  \citenamefont {Chen}, \citenamefont {Wang}, \citenamefont {Ma}, \citenamefont
  {Ma}, \citenamefont {Jin}, \citenamefont {Bai}, \citenamefont {Guimar\~aes},
  \citenamefont {Wang}, \citenamefont {Ma}, \citenamefont {Duan}, \citenamefont
  {Gao}, \citenamefont {Yu}, \citenamefont {Sun}, \citenamefont {Hu},
  \citenamefont {Xu}, \citenamefont {Wang}, \citenamefont {Yan}, \citenamefont
  {Zhou}, \citenamefont {Zhang}, \citenamefont {Zhou}, \citenamefont {Xu},
  \citenamefont {Xiao},\ and\ \citenamefont {Zhan}}]{Yang18}%
  \BibitemOpen
  \bibfield  {author} {\bibinfo {author} {\bibfnamefont {Y.~Y.}\ \bibnamefont
  {Yang}}, \bibinfo {author} {\bibfnamefont {X.}~\bibnamefont {Liu}}, \bibinfo
  {author} {\bibfnamefont {D.~Y.}\ \bibnamefont {Pang}}, \bibinfo {author}
  {\bibfnamefont {D.}~\bibnamefont {Patel}}, \bibinfo {author} {\bibfnamefont
  {R.~F.}\ \bibnamefont {Chen}}, \bibinfo {author} {\bibfnamefont {J.~S.}\
  \bibnamefont {Wang}}, \bibinfo {author} {\bibfnamefont {P.}~\bibnamefont
  {Ma}}, \bibinfo {author} {\bibfnamefont {J.~B.}\ \bibnamefont {Ma}}, \bibinfo
  {author} {\bibfnamefont {S.~L.}\ \bibnamefont {Jin}}, \bibinfo {author}
  {\bibfnamefont {Z.}~\bibnamefont {Bai}}, \bibinfo {author} {\bibfnamefont
  {V.}~\bibnamefont {Guimar\~aes}}, \bibinfo {author} {\bibfnamefont
  {Q.}~\bibnamefont {Wang}}, \bibinfo {author} {\bibfnamefont {W.~H.}\
  \bibnamefont {Ma}}, \bibinfo {author} {\bibfnamefont {F.~F.}\ \bibnamefont
  {Duan}}, \bibinfo {author} {\bibfnamefont {Z.~H.}\ \bibnamefont {Gao}},
  \bibinfo {author} {\bibfnamefont {Y.~C.}\ \bibnamefont {Yu}}, \bibinfo
  {author} {\bibfnamefont {Z.~Y.}\ \bibnamefont {Sun}}, \bibinfo {author}
  {\bibfnamefont {Z.~G.}\ \bibnamefont {Hu}}, \bibinfo {author} {\bibfnamefont
  {S.~W.}\ \bibnamefont {Xu}}, \bibinfo {author} {\bibfnamefont {S.~T.}\
  \bibnamefont {Wang}}, \bibinfo {author} {\bibfnamefont {D.}~\bibnamefont
  {Yan}}, \bibinfo {author} {\bibfnamefont {Y.}~\bibnamefont {Zhou}}, \bibinfo
  {author} {\bibfnamefont {Y.~H.}\ \bibnamefont {Zhang}}, \bibinfo {author}
  {\bibfnamefont {X.~H.}\ \bibnamefont {Zhou}}, \bibinfo {author}
  {\bibfnamefont {H.~S.}\ \bibnamefont {Xu}}, \bibinfo {author} {\bibfnamefont
  {G.~Q.}\ \bibnamefont {Xiao}}, \ and\ \bibinfo {author} {\bibfnamefont
  {W.~L.}\ \bibnamefont {Zhan}},\ }\href {\doibase 10.1103/PhysRevC.98.044608}
  {\bibfield  {journal} {\bibinfo  {journal} {Phys. Rev. C}\ }\textbf {\bibinfo
  {volume} {98}},\ \bibinfo {pages} {044608} (\bibinfo {year}
  {2018})}\BibitemShut {NoStop}%
\bibitem [{\citenamefont {Wang}\ \emph {et~al.}(2021)\citenamefont {Wang},
  \citenamefont {Yang}, \citenamefont {Moro}, \citenamefont {Guimar\~aes},
  \citenamefont {Lei}, \citenamefont {Pang}, \citenamefont {Duan},
  \citenamefont {Lou}, \citenamefont {Zamora}, \citenamefont {Wang},
  \citenamefont {Sun}, \citenamefont {Ong}, \citenamefont {Liu}, \citenamefont
  {Xu}, \citenamefont {Ma}, \citenamefont {Ma}, \citenamefont {Bai},
  \citenamefont {Hu}, \citenamefont {Xu}, \citenamefont {Gao}, \citenamefont
  {Yang}, \citenamefont {Jin}, \citenamefont {Zhang}, \citenamefont {Zhou},
  \citenamefont {Hu},\ and\ \citenamefont {Xu}}]{Wang21}%
  \BibitemOpen
  \bibfield  {author} {\bibinfo {author} {\bibfnamefont {K.}~\bibnamefont
  {Wang}}, \bibinfo {author} {\bibfnamefont {Y.~Y.}\ \bibnamefont {Yang}},
  \bibinfo {author} {\bibfnamefont {A.~M.}\ \bibnamefont {Moro}}, \bibinfo
  {author} {\bibfnamefont {V.}~\bibnamefont {Guimar\~aes}}, \bibinfo {author}
  {\bibfnamefont {J.}~\bibnamefont {Lei}}, \bibinfo {author} {\bibfnamefont
  {D.~Y.}\ \bibnamefont {Pang}}, \bibinfo {author} {\bibfnamefont {F.~F.}\
  \bibnamefont {Duan}}, \bibinfo {author} {\bibfnamefont {J.~L.}\ \bibnamefont
  {Lou}}, \bibinfo {author} {\bibfnamefont {J.~C.}\ \bibnamefont {Zamora}},
  \bibinfo {author} {\bibfnamefont {J.~S.}\ \bibnamefont {Wang}}, \bibinfo
  {author} {\bibfnamefont {Z.~Y.}\ \bibnamefont {Sun}}, \bibinfo {author}
  {\bibfnamefont {H.~J.}\ \bibnamefont {Ong}}, \bibinfo {author} {\bibfnamefont
  {X.}~\bibnamefont {Liu}}, \bibinfo {author} {\bibfnamefont {S.~W.}\
  \bibnamefont {Xu}}, \bibinfo {author} {\bibfnamefont {J.~B.}\ \bibnamefont
  {Ma}}, \bibinfo {author} {\bibfnamefont {P.}~\bibnamefont {Ma}}, \bibinfo
  {author} {\bibfnamefont {Z.}~\bibnamefont {Bai}}, \bibinfo {author}
  {\bibfnamefont {Q.}~\bibnamefont {Hu}}, \bibinfo {author} {\bibfnamefont
  {X.~X.}\ \bibnamefont {Xu}}, \bibinfo {author} {\bibfnamefont {Z.~H.}\
  \bibnamefont {Gao}}, \bibinfo {author} {\bibfnamefont {G.}~\bibnamefont
  {Yang}}, \bibinfo {author} {\bibfnamefont {S.~Y.}\ \bibnamefont {Jin}},
  \bibinfo {author} {\bibfnamefont {Y.~H.}\ \bibnamefont {Zhang}}, \bibinfo
  {author} {\bibfnamefont {X.~H.}\ \bibnamefont {Zhou}}, \bibinfo {author}
  {\bibfnamefont {Z.~G.}\ \bibnamefont {Hu}}, \ and\ \bibinfo {author}
  {\bibfnamefont {H.~S.}\ \bibnamefont {Xu}} (\bibinfo {collaboration} {RIBLL
  Collaboration}),\ }\href {\doibase 10.1103/PhysRevC.103.024606} {\bibfield
  {journal} {\bibinfo  {journal} {Phys. Rev. C}\ }\textbf {\bibinfo {volume}
  {103}},\ \bibinfo {pages} {024606} (\bibinfo {year} {2021})}\BibitemShut
  {NoStop}%
\bibitem [{\citenamefont {Jha}\ \emph {et~al.}(2020)\citenamefont {Jha},
  \citenamefont {Parkar},\ and\ \citenamefont {Kailas}}]{JHA20201}%
  \BibitemOpen
  \bibfield  {author} {\bibinfo {author} {\bibfnamefont {V.}~\bibnamefont
  {Jha}}, \bibinfo {author} {\bibfnamefont {V.}~\bibnamefont {Parkar}}, \ and\
  \bibinfo {author} {\bibfnamefont {S.}~\bibnamefont {Kailas}},\ }\href
  {\doibase https://doi.org/10.1016/j.physrep.2019.12.003} {\bibfield
  {journal} {\bibinfo  {journal} {Physics Reports}\ }\textbf {\bibinfo {volume}
  {845}},\ \bibinfo {pages} {1} (\bibinfo {year} {2020})},\ \bibinfo {note}
  {incomplete fusion reactions using strongly and weakly bound
  projectiles}\BibitemShut {NoStop}%
\bibitem [{\citenamefont {Ichimura}\ \emph {et~al.}(1985)\citenamefont
  {Ichimura}, \citenamefont {Austern},\ and\ \citenamefont {Vincent}}]{IAV85}%
  \BibitemOpen
  \bibfield  {author} {\bibinfo {author} {\bibfnamefont {M.}~\bibnamefont
  {Ichimura}}, \bibinfo {author} {\bibfnamefont {N.}~\bibnamefont {Austern}}, \
  and\ \bibinfo {author} {\bibfnamefont {C.~M.}\ \bibnamefont {Vincent}},\
  }\href {\doibase 10.1103/PhysRevC.32.431} {\bibfield  {journal} {\bibinfo
  {journal} {Phys. Rev. C}\ }\textbf {\bibinfo {volume} {32}},\ \bibinfo
  {pages} {431} (\bibinfo {year} {1985})}\BibitemShut {NoStop}%
\bibitem [{\citenamefont {Austern}\ \emph {et~al.}(1987)\citenamefont
  {Austern}, \citenamefont {Iseri}, \citenamefont {Kamimura}, \citenamefont
  {Kawai}, \citenamefont {Rawitscher},\ and\ \citenamefont
  {Yahiro}}]{Austern87}%
  \BibitemOpen
  \bibfield  {author} {\bibinfo {author} {\bibfnamefont {N.}~\bibnamefont
  {Austern}}, \bibinfo {author} {\bibfnamefont {Y.}~\bibnamefont {Iseri}},
  \bibinfo {author} {\bibfnamefont {M.}~\bibnamefont {Kamimura}}, \bibinfo
  {author} {\bibfnamefont {M.}~\bibnamefont {Kawai}}, \bibinfo {author}
  {\bibfnamefont {G.}~\bibnamefont {Rawitscher}}, \ and\ \bibinfo {author}
  {\bibfnamefont {M.}~\bibnamefont {Yahiro}},\ }\href {\doibase
  https://doi.org/10.1016/0370-1573(87)90094-9} {\bibfield  {journal} {\bibinfo
   {journal} {Physics Reports}\ }\textbf {\bibinfo {volume} {154}},\ \bibinfo
  {pages} {125 } (\bibinfo {year} {1987})}\BibitemShut {NoStop}%
\bibitem [{\citenamefont {Hussein}\ \emph {et~al.}(1990)\citenamefont
  {Hussein}, \citenamefont {Frederico},\ and\ \citenamefont
  {Mastroleo}}]{HUSSEIN1990269}%
  \BibitemOpen
  \bibfield  {author} {\bibinfo {author} {\bibfnamefont {M.}~\bibnamefont
  {Hussein}}, \bibinfo {author} {\bibfnamefont {T.}~\bibnamefont {Frederico}},
  \ and\ \bibinfo {author} {\bibfnamefont {R.}~\bibnamefont {Mastroleo}},\
  }\href {\doibase https://doi.org/10.1016/0375-9474(90)90159-J} {\bibfield
  {journal} {\bibinfo  {journal} {Nuclear Physics A}\ }\textbf {\bibinfo
  {volume} {511}},\ \bibinfo {pages} {269} (\bibinfo {year}
  {1990})}\BibitemShut {NoStop}%
\bibitem [{\citenamefont {Lei}\ and\ \citenamefont {{Moro}}(2015)}]{Jin15}%
  \BibitemOpen
  \bibfield  {author} {\bibinfo {author} {\bibfnamefont {J.}~\bibnamefont
  {Lei}}\ and\ \bibinfo {author} {\bibfnamefont {A.~M.}\ \bibnamefont
  {{Moro}}},\ }\href@noop {} {\bibfield  {journal} {\bibinfo  {journal} {Phys.
  Rev. C}\ }\textbf {\bibinfo {volume} {92}},\ \bibinfo {pages} {044616}
  (\bibinfo {year} {2015})}\BibitemShut {NoStop}%
\bibitem [{\citenamefont {Lei}\ and\ \citenamefont {Moro}(2015)}]{Jin15b}%
  \BibitemOpen
  \bibfield  {author} {\bibinfo {author} {\bibfnamefont {J.}~\bibnamefont
  {Lei}}\ and\ \bibinfo {author} {\bibfnamefont {A.~M.}\ \bibnamefont {Moro}},\
  }\href {\doibase 10.1103/PhysRevC.92.061602} {\bibfield  {journal} {\bibinfo
  {journal} {Phys. Rev. C}\ }\textbf {\bibinfo {volume} {92}},\ \bibinfo
  {pages} {061602} (\bibinfo {year} {2015})}\BibitemShut {NoStop}%
\bibitem [{\citenamefont {Lei}\ and\ \citenamefont {Moro}(2017)}]{Jin17}%
  \BibitemOpen
  \bibfield  {author} {\bibinfo {author} {\bibfnamefont {J.}~\bibnamefont
  {Lei}}\ and\ \bibinfo {author} {\bibfnamefont {A.~M.}\ \bibnamefont {Moro}},\
  }\href {\doibase 10.1103/PhysRevC.95.044605} {\bibfield  {journal} {\bibinfo
  {journal} {Phys. Rev. C}\ }\textbf {\bibinfo {volume} {95}},\ \bibinfo
  {pages} {044605} (\bibinfo {year} {2017})}\BibitemShut {NoStop}%
\bibitem [{\citenamefont {Lei}\ and\ \citenamefont {Moro}(2018)}]{Jin18}%
  \BibitemOpen
  \bibfield  {author} {\bibinfo {author} {\bibfnamefont {J.}~\bibnamefont
  {Lei}}\ and\ \bibinfo {author} {\bibfnamefont {A.~M.}\ \bibnamefont {Moro}},\
  }\href {\doibase 10.1103/PhysRevC.97.011601} {\bibfield  {journal} {\bibinfo
  {journal} {Phys. Rev. C}\ }\textbf {\bibinfo {volume} {97}},\ \bibinfo
  {pages} {011601} (\bibinfo {year} {2018})}\BibitemShut {NoStop}%
\bibitem [{\citenamefont {Lei}(2018)}]{Jin18b}%
  \BibitemOpen
  \bibfield  {author} {\bibinfo {author} {\bibfnamefont {J.}~\bibnamefont
  {Lei}},\ }\href {\doibase 10.1103/PhysRevC.97.034628} {\bibfield  {journal}
  {\bibinfo  {journal} {Phys. Rev. C}\ }\textbf {\bibinfo {volume} {97}},\
  \bibinfo {pages} {034628} (\bibinfo {year} {2018})}\BibitemShut {NoStop}%
\bibitem [{\citenamefont {Lei}\ and\ \citenamefont {Moro}(2019)}]{Jin19}%
  \BibitemOpen
  \bibfield  {author} {\bibinfo {author} {\bibfnamefont {J.}~\bibnamefont
  {Lei}}\ and\ \bibinfo {author} {\bibfnamefont {A.~M.}\ \bibnamefont {Moro}},\
  }\href {\doibase 10.1103/PhysRevLett.123.232501} {\bibfield  {journal}
  {\bibinfo  {journal} {Phys. Rev. Lett.}\ }\textbf {\bibinfo {volume} {123}},\
  \bibinfo {pages} {232501} (\bibinfo {year} {2019})}\BibitemShut {NoStop}%
\bibitem [{\citenamefont {Ferreira}\ \emph {et~al.}(2023)\citenamefont
  {Ferreira}, \citenamefont {Rangel}, \citenamefont {Lubian},\ and\
  \citenamefont {Canto}}]{Ferreira23}%
  \BibitemOpen
  \bibfield  {author} {\bibinfo {author} {\bibfnamefont {J.~L.}\ \bibnamefont
  {Ferreira}}, \bibinfo {author} {\bibfnamefont {J.}~\bibnamefont {Rangel}},
  \bibinfo {author} {\bibfnamefont {J.}~\bibnamefont {Lubian}}, \ and\ \bibinfo
  {author} {\bibfnamefont {L.~F.}\ \bibnamefont {Canto}},\ }\href {\doibase
  10.1103/PhysRevC.107.034603} {\bibfield  {journal} {\bibinfo  {journal}
  {Phys. Rev. C}\ }\textbf {\bibinfo {volume} {107}},\ \bibinfo {pages}
  {034603} (\bibinfo {year} {2023})}\BibitemShut {NoStop}%
\bibitem [{\citenamefont {Diaz-Torres}\ \emph {et~al.}(2007)\citenamefont
  {Diaz-Torres}, \citenamefont {Hinde}, \citenamefont {Tostevin}, \citenamefont
  {Dasgupta},\ and\ \citenamefont {Gasques}}]{Diaz07}%
  \BibitemOpen
  \bibfield  {author} {\bibinfo {author} {\bibfnamefont {A.}~\bibnamefont
  {Diaz-Torres}}, \bibinfo {author} {\bibfnamefont {D.~J.}\ \bibnamefont
  {Hinde}}, \bibinfo {author} {\bibfnamefont {J.~A.}\ \bibnamefont {Tostevin}},
  \bibinfo {author} {\bibfnamefont {M.}~\bibnamefont {Dasgupta}}, \ and\
  \bibinfo {author} {\bibfnamefont {L.~R.}\ \bibnamefont {Gasques}},\ }\href
  {\doibase 10.1103/PhysRevLett.98.152701} {\bibfield  {journal} {\bibinfo
  {journal} {Phys. Rev. Lett.}\ }\textbf {\bibinfo {volume} {98}},\ \bibinfo
  {pages} {152701} (\bibinfo {year} {2007})}\BibitemShut {NoStop}%
\bibitem [{\citenamefont {Marta}\ \emph {et~al.}(2014)\citenamefont {Marta},
  \citenamefont {Canto},\ and\ \citenamefont {Donangelo}}]{Marta14}%
  \BibitemOpen
  \bibfield  {author} {\bibinfo {author} {\bibfnamefont {H.~D.}\ \bibnamefont
  {Marta}}, \bibinfo {author} {\bibfnamefont {L.~F.}\ \bibnamefont {Canto}}, \
  and\ \bibinfo {author} {\bibfnamefont {R.}~\bibnamefont {Donangelo}},\ }\href
  {\doibase 10.1103/PhysRevC.89.034625} {\bibfield  {journal} {\bibinfo
  {journal} {Phys. Rev. C}\ }\textbf {\bibinfo {volume} {89}},\ \bibinfo
  {pages} {034625} (\bibinfo {year} {2014})}\BibitemShut {NoStop}%
\bibitem [{\citenamefont {Kolinger}\ \emph {et~al.}(2018)\citenamefont
  {Kolinger}, \citenamefont {Canto}, \citenamefont {Donangelo},\ and\
  \citenamefont {Souza}}]{Kolinger18}%
  \BibitemOpen
  \bibfield  {author} {\bibinfo {author} {\bibfnamefont {G.~D.}\ \bibnamefont
  {Kolinger}}, \bibinfo {author} {\bibfnamefont {L.~F.}\ \bibnamefont {Canto}},
  \bibinfo {author} {\bibfnamefont {R.}~\bibnamefont {Donangelo}}, \ and\
  \bibinfo {author} {\bibfnamefont {S.~R.}\ \bibnamefont {Souza}},\ }\href
  {\doibase 10.1103/PhysRevC.98.044604} {\bibfield  {journal} {\bibinfo
  {journal} {Phys. Rev. C}\ }\textbf {\bibinfo {volume} {98}},\ \bibinfo
  {pages} {044604} (\bibinfo {year} {2018})}\BibitemShut {NoStop}%
\bibitem [{\citenamefont {Cook}\ \emph {et~al.}(2019)\citenamefont {Cook},
  \citenamefont {Simpson}, \citenamefont {Bezzina}, \citenamefont {Dasgupta},
  \citenamefont {Hinde}, \citenamefont {Banerjee}, \citenamefont {Berriman},\
  and\ \citenamefont {Sengupta}}]{Cook19}%
  \BibitemOpen
  \bibfield  {author} {\bibinfo {author} {\bibfnamefont {K.~J.}\ \bibnamefont
  {Cook}}, \bibinfo {author} {\bibfnamefont {E.~C.}\ \bibnamefont {Simpson}},
  \bibinfo {author} {\bibfnamefont {L.~T.}\ \bibnamefont {Bezzina}}, \bibinfo
  {author} {\bibfnamefont {M.}~\bibnamefont {Dasgupta}}, \bibinfo {author}
  {\bibfnamefont {D.~J.}\ \bibnamefont {Hinde}}, \bibinfo {author}
  {\bibfnamefont {K.}~\bibnamefont {Banerjee}}, \bibinfo {author}
  {\bibfnamefont {A.~C.}\ \bibnamefont {Berriman}}, \ and\ \bibinfo {author}
  {\bibfnamefont {C.}~\bibnamefont {Sengupta}},\ }\href {\doibase
  10.1103/PhysRevLett.122.102501} {\bibfield  {journal} {\bibinfo  {journal}
  {Phys. Rev. Lett.}\ }\textbf {\bibinfo {volume} {122}},\ \bibinfo {pages}
  {102501} (\bibinfo {year} {2019})}\BibitemShut {NoStop}%
\bibitem [{\citenamefont {Rawitscher}(1974)}]{Raw74}%
  \BibitemOpen
  \bibfield  {author} {\bibinfo {author} {\bibfnamefont {G.~H.}\ \bibnamefont
  {Rawitscher}},\ }\href {\doibase 10.1103/PhysRevC.9.2210} {\bibfield
  {journal} {\bibinfo  {journal} {Phys. Rev. C}\ }\textbf {\bibinfo {volume}
  {9}},\ \bibinfo {pages} {2210} (\bibinfo {year} {1974})}\BibitemShut
  {NoStop}%
\bibitem [{\citenamefont {Yahiro}\ \emph {et~al.}(1986)\citenamefont {Yahiro},
  \citenamefont {Iseri}, \citenamefont {Kameyama}, \citenamefont {Kamimura},\
  and\ \citenamefont {Kawai}}]{Yah86}%
  \BibitemOpen
  \bibfield  {author} {\bibinfo {author} {\bibfnamefont {M.}~\bibnamefont
  {Yahiro}}, \bibinfo {author} {\bibfnamefont {Y.}~\bibnamefont {Iseri}},
  \bibinfo {author} {\bibfnamefont {H.}~\bibnamefont {Kameyama}}, \bibinfo
  {author} {\bibfnamefont {M.}~\bibnamefont {Kamimura}}, \ and\ \bibinfo
  {author} {\bibfnamefont {M.}~\bibnamefont {Kawai}},\ }\href@noop {}
  {\bibfield  {journal} {\bibinfo  {journal} {Progress of Theoretical Physics
  Supplement}\ }\textbf {\bibinfo {volume} {89}},\ \bibinfo {pages} {32}
  (\bibinfo {year} {1986})}\BibitemShut {NoStop}%
\bibitem [{\citenamefont {Austern}\ \emph {et~al.}(1989)\citenamefont
  {Austern}, \citenamefont {Yahiro},\ and\ \citenamefont {Kawai}}]{Austern89}%
  \BibitemOpen
  \bibfield  {author} {\bibinfo {author} {\bibfnamefont {N.}~\bibnamefont
  {Austern}}, \bibinfo {author} {\bibfnamefont {M.}~\bibnamefont {Yahiro}}, \
  and\ \bibinfo {author} {\bibfnamefont {M.}~\bibnamefont {Kawai}},\ }\href
  {\doibase 10.1103/PhysRevLett.63.2649} {\bibfield  {journal} {\bibinfo
  {journal} {Phys. Rev. Lett.}\ }\textbf {\bibinfo {volume} {63}},\ \bibinfo
  {pages} {2649} (\bibinfo {year} {1989})}\BibitemShut {NoStop}%
\bibitem [{\citenamefont {Faddeev}(1961)}]{Faddeev}%
  \BibitemOpen
  \bibfield  {author} {\bibinfo {author} {\bibfnamefont {L.~D.}\ \bibnamefont
  {Faddeev}},\ }\href@noop {} {\bibfield  {journal} {\bibinfo  {journal} {Sov.
  Phys. JETP}\ }\textbf {\bibinfo {volume} {12}},\ \bibinfo {pages} {1014}
  (\bibinfo {year} {1961})},\ \bibinfo {note} {[Zh. Eksp. Teor.
  Fiz.39,1459(1960)]}\BibitemShut {NoStop}%
%%CITATION = SPHJA,12,1014;%%
\bibitem [{\citenamefont {Thompson}\ and\ \citenamefont
  {Nunes}(2009)}]{thompson_nunes_2009}%
  \BibitemOpen
  \bibfield  {author} {\bibinfo {author} {\bibfnamefont {I.~J.}\ \bibnamefont
  {Thompson}}\ and\ \bibinfo {author} {\bibfnamefont {F.~M.}\ \bibnamefont
  {Nunes}},\ }\href {\doibase 10.1017/CBO9781139152150} {\emph {\bibinfo
  {title} {Nuclear Reactions for Astrophysics: Principles, Calculation and
  Applications of Low-Energy Reactions}}}\ (\bibinfo  {publisher} {Cambridge
  University Press},\ \bibinfo {year} {2009})\BibitemShut {NoStop}%
\bibitem [{\citenamefont {Baylis}\ and\ \citenamefont
  {Peel}(1982)}]{BAYLIS19827}%
  \BibitemOpen
  \bibfield  {author} {\bibinfo {author} {\bibfnamefont {W.}~\bibnamefont
  {Baylis}}\ and\ \bibinfo {author} {\bibfnamefont {S.}~\bibnamefont {Peel}},\
  }\href {\doibase https://doi.org/10.1016/0010-4655(82)90039-X} {\bibfield
  {journal} {\bibinfo  {journal} {Computer Physics Communications}\ }\textbf
  {\bibinfo {volume} {25}},\ \bibinfo {pages} {7 } (\bibinfo {year}
  {1982})}\BibitemShut {NoStop}%
\bibitem [{\citenamefont {Druet}\ \emph {et~al.}(2010)\citenamefont {Druet},
  \citenamefont {Baye}, \citenamefont {Descouvemont},\ and\ \citenamefont
  {Sparenberg}}]{DRUET201088}%
  \BibitemOpen
  \bibfield  {author} {\bibinfo {author} {\bibfnamefont {T.}~\bibnamefont
  {Druet}}, \bibinfo {author} {\bibfnamefont {D.}~\bibnamefont {Baye}},
  \bibinfo {author} {\bibfnamefont {P.}~\bibnamefont {Descouvemont}}, \ and\
  \bibinfo {author} {\bibfnamefont {J.-M.}\ \bibnamefont {Sparenberg}},\ }\href
  {\doibase https://doi.org/10.1016/j.nuclphysa.2010.05.060} {\bibfield
  {journal} {\bibinfo  {journal} {Nuclear Physics A}\ }\textbf {\bibinfo
  {volume} {845}},\ \bibinfo {pages} {88} (\bibinfo {year} {2010})}\BibitemShut
  {NoStop}%
\bibitem [{\citenamefont {Kasano}\ and\ \citenamefont
  {Ichimura}(1982)}]{Kas82}%
  \BibitemOpen
  \bibfield  {author} {\bibinfo {author} {\bibfnamefont {A.}~\bibnamefont
  {Kasano}}\ and\ \bibinfo {author} {\bibfnamefont {M.}~\bibnamefont
  {Ichimura}},\ }\href@noop {} {\bibfield  {journal} {\bibinfo  {journal}
  {Physics Letters B}\ }\textbf {\bibinfo {volume} {115}},\ \bibinfo {pages}
  {81} (\bibinfo {year} {1982})}\BibitemShut {NoStop}%
\bibitem [{\citenamefont {Han}\ \emph {et~al.}(2006)\citenamefont {Han},
  \citenamefont {Shi},\ and\ \citenamefont {Shen}}]{Han06}%
  \BibitemOpen
  \bibfield  {author} {\bibinfo {author} {\bibfnamefont {Y.}~\bibnamefont
  {Han}}, \bibinfo {author} {\bibfnamefont {Y.}~\bibnamefont {Shi}}, \ and\
  \bibinfo {author} {\bibfnamefont {Q.}~\bibnamefont {Shen}},\ }\href {\doibase
  10.1103/PhysRevC.74.044615} {\bibfield  {journal} {\bibinfo  {journal} {Phys.
  Rev. C}\ }\textbf {\bibinfo {volume} {74}},\ \bibinfo {pages} {044615}
  (\bibinfo {year} {2006})}\BibitemShut {NoStop}%
\bibitem [{\citenamefont {Watanabe}(1958)}]{Watanabe58}%
  \BibitemOpen
  \bibfield  {author} {\bibinfo {author} {\bibfnamefont {S.}~\bibnamefont
  {Watanabe}},\ }\href {\doibase https://doi.org/10.1016/0029-5582(58)90180-9}
  {\bibfield  {journal} {\bibinfo  {journal} {Nuclear Physics}\ }\textbf
  {\bibinfo {volume} {8}},\ \bibinfo {pages} {484} (\bibinfo {year}
  {1958})}\BibitemShut {NoStop}%
\end{thebibliography}%
\end{document}